\documentstyle[preprint, eqsecnum, aps]{revtex}

\tightenlines
\begin{document}

\preprint{December 27, 2001}

\bigskip\bigskip\bigskip\bigskip
\title{On  Variations\\
 in Discrete Mechanics and Field Theory
}


\author{Han-Ying Guo
}
\address
{Institute of Theoretical Physics, Chinese Academia of Sciences, \\[0pt]
P.O. Box 2735, Beijing 100080, China}
\author{
Ke Wu
}
\address
{Department of Mathematics, Capital Normal University,\\[0pt]
 Beijing 100037, China\\[2mm]
Institute of Theoretical Physics, Chinese Academia of Sciences, \\[0pt]
P.O. Box 2735, Beijing 100080, China  }


\vspace {1.5cm}

\maketitle
\begin{abstract}

Some problems on variations  are raised for classical discrete
mechanics and field theory and the difference variational approach
with variable step-length is proposed motivated by Lee's approach
to discrete mechanics and the difference discrete variational
principle for difference discrete mechanics and field theory on
regular lattice. Based upon Hamilton's principle for the vertical
variations and double operation of vertical exterior differential
on action, it is shown that for both continuous and variable
step-length difference cases there exists the nontrivial
Euler-Lagrange cohomology as well as the necessary and sufficient
condition for symplectic/multi-symplectic structure preserving
properties is the relevant Euler-Lagrange 1-form is closed in both
continuous and difference classical mechanics and field theory.
While the horizontal variations give rise to the relevant
identities or relations of the Euler-Lagrange equation and
conservation law of the energy/energy-momentum tensor for
continuous or discrete systems. The total variations are also
discussed. Especially, for those discrete cases the variable
step-length of the  difference is determined by the relation
between the Euler-Lagrange equation and conservation law of the
energy/energy-momentum tensor. In addition, this approach together
with difference version of the Euler-Lagrange cohomology can be
applied not only to discrete Lagrangian formalism but also to the
Hamiltonian formalism for difference mechanics and field theory.
\end{abstract}


\vspace {0.5cm}

\section{Introduction}

Variation problems play a fundamental, even central, role in
various types of continuous mechanics (see, for example,
\cite{c1}, \cite{ald78}, \cite{am78}) and field theories for both
Lagrangian and Hamiltonian formalisms, which can be transformed
each other in many cases via Legendre transformation. A lot of
important issues such as the equations of motions, the (intrinsic)
symplectic or multisymplectic preserving properties, conservation
laws associated with certain symmetries, topological properties
etc. are very closely related to the variation problems with fixed
or variable domains. On the other hand, however, there are  some
open questions relevant to variation problems in different kinds
of discrete mechanics, field theories as well as corresponding
symplectic and multisymplectic algorithms, although discrete
variation problems still play an important role, particularly, for
the discrete Lagrangian formalism of these discrete systems. In
fact, for a long period, there had been no any discrete variation
approach available to both discrete Lagrangian and Hamiltonian
formalisms that relate each other by discrete Legendre
transformation until the difference discrete variational approach
has been proposed very recently \cite{glw}, \cite{glww}.

As far as discretized version with difference for mechanics is
concerned, Lee suggested a discrete variational approach to
discrete Lagrangian mechanics and relevant algorithm in early
1980's \cite{td1}, \cite{td2}, \cite{td3}. In Lee's approach the
time steps are variable in view of time being a dynamical variable
in order to preserve the energy discretely. Veselov \cite{av88},
\cite{MV91} also proposed the discrete variational principle by
the end of 1980's, which is almost the same as Lee's approach
except without taking variation with respect to the discrete time
so that it does not keep the conservation of the energy discretely
in general. In addition, both approaches are merely available to
discrete Lagrangian mechanics and nothing to do with discrete
Hamiltonian mechanics.

On the other hand,
 Ruth \cite{ruth83} and Feng \cite{fk84} proposed the symplectic
algorithm for Hamiltonian mechanics. In this algorithm (for a
review, see \cite{sc94}), the time step-length is fixed and the
symplectic preserving property is discretely kept. However, the
discrete version of energy conservation can not be maintained
discretely in general. The symplectic algorithm  plays very
important role in computational mathematics and its applications
cover various branches in sciences. It is well known that the
progresses of symplectic algorithm promote further development of
the structure-preserving algorithms. In these algorithms, the time
step-length is always being fixed so that the price paid by
keeping structure preserving is the loss of other conservation
laws for the continuous cases in general.

{\bf Structure preserving criterion}: It is widely accepted that
the discrete systems should be thought as the discrete
counterparts of the corresponding continuous systems. However, in
order to discretize continuous systems, a guide line in needed. In
Feng's first paper on the symplectic algorithm, he wrote a working
hypothesis: ``It is natural to look forward to those discrete
systems which preserve as much as possible the intrinsic
properties of the continuous system.'' \cite{fk84} In fact, this
statement should be regarded as a criterion, the
structure-preserving criterion, for constructing mostly quarried
one in all kinds of structure-preserving algorithms.  However, in
order to carry through this criterion, it is needed to know how to
answer the following simple questions.

{\bf Problem 1}: What are the most important intrinsic
``structures" in continuous systems, such as classical mechanics
and field theory, that should also be maintained in the course of
discretization? What are the discrete counterparts of these
``structures" and how to preserve them in certain discrete
version? What about the lowest price has to be paid in the course
of discretization?

It is well known that there are two classes of conservation laws
in canonical conservative mechanics. The first class of
conservation laws is that of phase-area conservation laws
characterized by the symplectic preserving property and another
class is related to energy and all first integrals of the
canonical equations. Thus, the following questions can apparently
be raised.

{\bf Problem 2}: Is it possible to establish a kind of discrete
mechanics and/or  structure-preserving algorithms in such a way
that they not only discretely preserve the symplectic property but
also some other conservation laws, specially the energy
conservation? Do these discrete systems can be established by a
discrete variational approach? Does this discrete variational
approach can be applied to both discrete Lagrangian and
Hamiltonian formalisms?

Although in Lee's approach, it should be able to prove that in
addition to discrete energy conservation the symplectic structure
is also preserved since Lee's approach is a discrete variational
approach \footnote{In \cite{kmo99} this problem has been partly
solved by defining a conserved discrete energy in Veselov's
approach. The complete resolving to this problem in Lee's
framework has been made until very recently in \cite{cgw},
\cite{cgw1}.}. But the framework of either Lee's approach or
Veselov's one can not be applied to the discrete Hamiltonian
systems. To our knowledge, therefore, these problems are still
open, at least partly.

In the cases of discrete field theories and multisymplectic
algorithms, the symplectic algorithm has been generalized to the
multisymplectic one in what is called ``Hamiltonian formalism"
\cite{TB97}. On the other hand, Veselov's discrete variation
approach to the discrete mechanics has also been generalized to
field theories in Lagrangian formalism to get so-called ``
variational multisymplectic integrators" \cite{MS98},
\cite{MPS98}. In both approaches, the step-lengths are fixed so
that the energy-momentum tensor cannot be conserved in general
although the multisymplectic structure preserving property in
field theory can be maintained discretely in certain manner. Thus
a set of similar questions can also be raised to the discrete
field theory and multisymplectic algorithms.

{\bf Problem 3}:  Is it possible to establish a  discrete
variational approach to describe a kind of discrete field theories
and/or multisymplectic algorithms in such a way that not only the
multisymplectic property is discretely  preserved but also the
conservation laws such as energy-momentum conservation law are
discretely maintained in certain version? Is it possible to apply
such an approach in both discrete Lagrangian and Hamiltonian
formalisms? To our knowledge, these problems are still open as
well.

As was just mentioned, for the difference discrete variational
approach to discrete mechanics and field theory with fixed
step-length differences, it can be applied to both Lagrangian and
Hamiltonian formalisms \cite{glw}, \cite{glww}. The key point of
this approach is to regard the differences with fixed step-length
as a kind of entire geometric objects, which play an analogical
role with the one played by derivative in continuous cases.

{\bf Problem 4}: Is it possible to generalize the difference
discrete variational approach with fixed step-length differences
to the one  with varied step-length differences so as to the
discrete energy conservation law in certain version may be kept
together with the symplectic and/or multisymplectic preserving
properties?

In order to carry through the structure-preserving criterion, we
present an approach to the discrete variation problems. Namely,
the difference variational approach with variable step-lengths
named variable difference variational approach to these problems
in discrete mechanics and field theory. In other wards, this
approach is available to the discrete total variations with
keeping the step-length of differences to be varied by an equation
from the variation of the discrete action with respect to discrete
time and/or discrete space coordinates. As was mentioned, this
approach is a natural generalization of the difference variational
approach with fixed step-lengths proposed recently in \cite{glw},
\cite{glww} for the ordinary discrete variation problems with
fixed discrete domain. In fact, the approach in \cite{glw},
\cite{glww} is just discrete vertical variation so that it is
natural to keep the step-length being fixed. The most important
key point of this approach is that in the course of calculation of
variation problems in discrete mechanics and field theory, the
differences with variable step-lengths are kept as entire
geometric objects as much as possible. Consequently, this discrete
variation approach not only keep the advantage of Lee's discrete
variation, which conserves the energy of the system discretely,
but also the advantage in variation in Veselov type, which is
symplectic or multisymplectic. In addition, this variable
difference  variation approach can be applied not only to the
Lagrangian formalism but the Hamiltonian formalism as well for
both discrete mechanics and discrete field theory, since the
discrete canonical `` momenta" and discrete version of Legendre
transformation can be introduced in terms of variable step-length
differences. For simplicity, we consider in this paper the
discrete Lagrangian of first order of difference only.

The paper is organized as follows. In section 2, we recall the
total variation problems in Lagrangian and Hamiltonian mechanics.
In section 3,
 we present the variable difference variational
 approach and deal with the difference Lagrangian and Hamiltonian
 mechanics. In section 4, we recall briefly the total variation
 problems in Lagrangian and Hamiltonian field theory with generic variables.
 In section 5, we apply the variable difference discrete variational
 approach to the total discrete variation
 problems in difference discrete Lagrangian and Hamiltonian field theory
 with generic variables. Finally, we end with some remarks.

\section{General Variations for Classical Mechanics}

Let us recall briefly the general or total variation
calculus with variable domain in classical mechanics.

Let  $t \in T \simeq R$ be the time, $M$ an $n$-dimensional
configuration space. Consider a fibre bundle $E(T, Q, \pi)$ 
with projection $\pi: E \rightarrow T$ on $T$,
${\pi}^{-1}: t\rightarrow Q_t$ isomorphic to $M$ is the fibre on
$t \in T$. Denote $\Gamma(E)$ the sections on $E$, $TE$ the
tangent bundle of $E$, $T_v E\subset TE$ the vertical sub-bundle
of $TE$, etc..

\subsection{Variations in Lagrangian mechanics}

We first consider the Lagrangian mechanics. The Lagrangian  of the
system is denoted as $L(q^i(t), \dot q^i(t); t)$, $i=1,\cdots ,n$,
which is a mapping from $TE$ 
to $R$. For
simplicity, the Lagrangian is of the first order. The action
functional  is
\begin{equation}
{S}([q^i(t)]; t_1, t_2)=\int_{t_1}^{t_2}dt {L}(q^i(t), \dot
q^i(t); t). \label{actn}
\end{equation}
Here $q^i$'s are coordinates in the fiber, $q^i(t)$ describes a
curve ${C}_a^b$ with ending points $a$ and $b$, $t_a=t_1,
t_b=t_2$, along which the motion of the system is assumed to take
place, and $\dot q^i(t)=dq^i(t)/dt$.

Let us consider the general variation of $q^i(t)$
\begin{equation}\label{vrtnq}
q^i(t) \rightarrow q'^i(t')=q^i(t)+\delta_tq^i(t)%
\end{equation}
accompanied with  an infinitesimal
re-parameterization of time $t$ 
\begin{equation}\label{vrtnt}
 t \rightarrow t'(t)=t+\delta t.%
\end{equation}
Here $\delta_tq^i(t)$ denotes the total variation that can be
divided into two parts,
\begin{equation}\begin{array}{rcl}\label{vrtq}
\delta_tq^i(t)&=&\delta_v q^i(t)+\delta_h q^i(t),\\
\delta_v q^i(t)&=&q'^i(t)-q^i(t),\\
 \delta_h
q^i(t)&=&q'^i(t')-q'^i(t)\\
&=&q^i(t')-q^i(t)+O(\delta^2)=\delta
t\frac{d}{dt}q^i(t)+O(\delta^2),
\end{array}\end{equation} where $\delta_v q^i(t)$ denotes the equal time
part variation or the vertical one and $\delta_h q^i(t)$ the
horizontal part  along the fibre induced by  the
re-parameterization of the time $t$ (\ref{vrtnt}). If we introduce
a variational vector field on $T$
\begin{equation}\label{vvft}
\xi(t):=\delta t \frac{d}{dt},
\end{equation}
the horizontal variation $\delta_h q^i(t)$ is the Lie derivative
of $q^i(t)$ with respect to the variation vector field
(\ref{vvft}).

Similarly,
\begin{equation}\begin{array}{rcl}\label{vrtdotq}
\delta_t\dot q^i(t)&=&(\delta_v+\delta_h)\dot q^i(t),\\
\delta_v \dot q^i(t)&=&\dot q'^i(t)-\dot q^i(t),\\
\delta_h \dot q^i(t)&=&\frac{d}{dt'} q'^i(t')-\frac{d}{dt'} q'^i(t)\\
&=&\frac{d}{dt'} q^i(t')-\frac{d}{dt} q^i(t)+O(\delta^2)=\delta t
\frac{d}{dt}\dot q^i(t)
+O(\delta^2).
\end{array}\end{equation}
Note that $\delta_h \dot q^i(t)$ is also the Lie derivative of
$\dot q^i(t)$ with respect to the variational vector field
(\ref{vvft}). In fact, this is true for  a kind of functionals of
$q^i(t)$, $\dot
q^i(t)$ and $t$: 
\begin{equation}\label{lied1}
\delta_h F(q^i(t), \dot q^i(t), t)={L}_\xi F(q^i(t), \dot q^i(t), t).
\end{equation}
To the time change  (\ref{vrtnt}) is associated the change in the
measure in (\ref{actn}) given by the Jacobi formula
\begin{equation}
dt'=\frac{\partial t'}{\partial t}dt= (1+\frac{d}{dt}\delta t)dt,
\end{equation}%
i.e.
\begin{equation}%
 \delta(dt)=d(t+\delta t)-dt=dt\frac{d}{dt}\delta t.
\end{equation}
It is easy to see that this change in the measure is also the Lie
derivative of the measure with respect to the variation vector
(\ref{vvft}):
$$
\delta(dt)={L}_\xi dt=d {L}_\xi t=d \delta t.%
$$
Here the commuting property between $d$ on $T$ and the Lie
derivative is used.

Now the Lagrangian is changed as
\begin{equation}\label{ldfm}
{L}(q^i(t), \frac{d}{d t} q^i(t); t)\rightarrow{L}(q'^i(t'),
\frac{d}{d t'} q'^i(t'); t')={L}(q^i(t),
\frac{d}{d t} q^i(t); t)+\delta_t{L}, 
\end{equation}
and the action is also deformed to
\begin{equation}\begin{array}{crl}\label{actn2}
 {S}([ q'^i(t')]; t'_1, t'_2)&=&\int_{t'_1}^{t'_2}dt' {L}(q'^i(t'),
\frac{d}{d t'} q'^i(t'); t')\\
&=&\int_{t_1}^{t_2} \frac{\partial t'}{\partial t}dt \{{
L}(q^i(t), \dot q^i(t); t)+\delta_t{L}\}\\
&=&\int_{t_1}^{t_2}dt \{{L}+(\frac{d}{dt}\delta t) {
L}+\delta_t{L}\}\\
&=&{S}([ q^i(t)]; t_1, t_2)+\delta_t{S}%
\end{array}\end{equation}
A more or less straightforward calculation shows
\begin{equation}\label{vrts}
\delta_t{S}=\int_{t_1}^{t_2}dt \{[{L}_{q^i}] \delta_tq^i +[
\frac{d}{dt} {H} +\frac{\partial}{\partial t}{L}]\delta t
+\frac{d}{dt}(\frac{\partial {L}}{\partial \dot q^j}\delta_t
q^j-{H}\delta t)\},%
\end{equation}
where $[{L}_{q^i}]$ is the Euler-Lagrange operator and $H$ is the energy
(Hamiltonian)%
\begin{equation}
[{L}_{q^i}]:=\frac{\partial {L}}{\partial
q^i}-\frac{d}{dt}(\frac{\partial {L}}{\partial \dot q^i}),\qquad
{H}:=\frac{\partial {L}}{\partial \dot q^i}\dot q^i-{L}.%
\end{equation}

The vertical and horizontal variations should be  separated
as the independent ones and this leads to
\begin{equation}\label{vvsm}
\delta_v {S}=\int_{t_1}^{t_2}dt \{[{L}_{q^ i}] \delta_v q^i
+\frac{d}{dt}(\frac{\partial {L}}{\partial \dot q^j}\delta_v
q^j)\},
\end{equation}
and
\begin{equation}
\delta_h {S}=\int_{t_1}^{t_2}dt \{[{L}_{q^i}] \delta_h q^i +[
\frac{d}{dt} {H} +\frac{\partial}{\partial t}{L}]\delta t
+\frac{d}{dt}(\frac{\partial {L}}{\partial \dot q^j}\delta_h
q^j-{H}\delta t)\}.
\end{equation}
For the vertical part, the Hamilton's principle leads to the
Euler-Lagrange equation  if $\delta_v q^j|_{t_1}=\delta_v
q^j|_{t_2}=0$
\begin{equation}\label{eleqm}
\frac{\partial {L}}{\partial q^i}-\frac{d}{dt}(\frac{\partial
{L}}{\partial \dot q^i})=0.
\end{equation}

For the horizontal part, however, it is easy to check
\begin{equation}\label{eqclw}
\begin{array}{l}
[{L}_{q^ i}] \dot q^i+\frac{d}{dt}{ H}+\frac{\partial}{\partial
t}{L}=0,\\
\vartheta :=\frac{\partial {L}}{\partial \dot q^j}\delta_h
q^j-{H}\delta t={L}\delta t.
\end{array}
\end{equation}
Therefore,
\begin{equation}\label{vrtnhs}
\delta_h {S}=\int_{t_1}^{t_2}dt \frac{d}{dt}({L} \delta t)=0.
\end{equation}
This is just the invariance of the action $S$ with respect to the
re-parameterization of time. Of course, from the first equation of
(\ref{eqclw}), it still follows  the conservation law for the
energy   if and only if the Euler-Lagrange equation is satisfied
and $L$ does not depend on $t$ manifestly.

 If  the identities in (\ref{eqclw}) is  employed directly, it follows
(see, for example, \cite{c1}) that
\begin{equation}\label{vrtactn}
\delta_t{S}=\int_{t_1}^{t_2}dt \{[{L}_{q^ i}] \delta_v q^i
+\frac{d}{dt}(\frac{\partial {L}}{\partial \dot q^j}\delta_v q^j+
{L}\delta t)\}.
\end{equation}
A quantity now can be defined
\begin{equation}\label{crtm}
{J}:=\frac{\partial {L}} {\partial \dot q^j}\delta_v q^j+
{L}\delta t,
\end{equation}
the invariance of ${S}$ under the re-parameterization of $t$, i.e.
$\delta_t{S}=0$, leads to the conservation of the quantity ${ J}$
if and only if the Euler-Lagrange equation is satisfied:
\begin{equation}\label{csvm}
\frac{d}{dt}{J}=0.
\end{equation}

\vskip 3mm
 {\it\bf Remark 2.1:}

 Introducing an exterior differential operator
 $d_v$ along the fibre that
satisfies
\begin{equation}\label{dv}
{d_v}^2=0, \qquad \{d_v, d_h\}=0, \qquad d:=d_v+d_h,
\end{equation}
where $d_h$ and $d$ is the nilpotent exterior differential
operator on $T^*{ T}$ and $T^*E$ respectively as was as a vertical
variational vector field
\begin{equation}\label{vvvfq}
\xi_q:=\delta_v q^i(t) \frac{\partial}{\partial q^i},
\end{equation}
then
\begin{equation}\label{vvq}
\delta_v q^i(t) =i_{\xi_q}d_vq^i=i_{\xi_q}dq^i.
\end{equation}

By means of the vertical variational vector field (\ref{vvvfq}) on
$TQ$, $\delta_v{S}$ can also be expressed as its contraction with
1-form $d_v{S} \in T^* Q $ 
\begin{equation}\label{vrs}
i_{\xi_q} d_v{ S} =\delta_v{S}.
\end{equation}
We may calculate $d_v{ S} \in T^* Q$. Since $d_v$ commutes with
the integral of $dt$ (see also, for example, the functional
differential calculus in \cite{olv93}), it is straightforward to
get
\begin{equation}\label{dvs1}
 d_v{ S}=\int_{t_1}^{t_2}dt \{[{ L}_{q^i}] d_v q^i
+\frac{d}{dt}\theta
)\},%
\end{equation}
where $\theta$ is the Lagrange
1-form%
\begin{equation}
\theta:=\frac{\partial { L}}{\partial \dot q^i}d_v q^i.%
\end{equation}
Now by contracting with the vertical variational vector field
(\ref{vvvfq}) it follows straightforwardly  $\delta_v S$ in
(\ref{vvsm}).

Furthermore, due to the nilpotency of $d_v$, it is easy to
get\begin{equation}\label{dE}
d_v{\cal E}+\frac{d}{dt}\omega=0,
\end{equation}
where $\cal E$ is called the Euler-Lagrange 1-form \cite{glw},
\cite{glww1},\cite{glww}, defined by
\begin{equation}\label{elfm}
{\cal E}(q^i(t), \dot q^i(t); t):=[{ L}_{q^i}] d_v q^i,
\end{equation}
$\omega$ is the symplectic 2-form and in local
coordinates:
\begin{equation}\label{defspm}
\omega:=d_v \theta=\frac {\partial^2 { L}} {{\partial q^j}
\partial {\dot q^i} } d_v q^j \wedge d_v q^i +\frac {\partial^2 { L}}
{\partial {\dot q^j}{\partial {\dot q^i}}} d_v{\dot q}^j \wedge
d_v q^i.
\end{equation}
From (\ref{elfm}), (\ref{dvs1}) and (\ref{dE}), the following
theorem can be proved \cite{glw}, \cite{glww1}, \cite{glww}:

\vskip 3mm {\it Theorem 1: For all Lagrangian of a kind of systems
with first order of derivatives on the bundle $E(T, Q, \pi)$

1. The following Euler-Lagrange cohomology is  nontrivial:
$$H_{LM}:=\{{\cal E}|d_v{\cal E}=0\}/ \{{\cal E}|{\cal
E}=d_v \alpha\},$$
where $\alpha=\alpha(q^i(t), \dot q^i(t);
t)$ is an arbitrary function of $(q^i(t), \dot q^i(t); t)$.

2. The necessary and sufficient condition for conservation of the
symplectic 2-form, i.e.
\begin{equation}\label{csvsy}
\frac{d}{dt}\omega=0,
\end{equation}
is the corresponding Euler-Lagrange 1-form being closed. }

\vskip 3mm
 {\it\bf Remark 2.2:}

  From the definition of the Lie
derivative it can be seen that the horizontal variations are given
by the Lie derivative with respect to the variational vector
field.

Let $\xi$ be a vector field on $T$, $exp(\lambda \xi)$ be the flow
with parameter $\lambda$, i.e. the one-parameter diffeomorphism
group, induced by $\xi$, $f$ a differential or a vector on $T$.
The infinitesimal change of $f$ under flow is described by its Lie
derivative  with respect to the vector field $\xi$
\begin{equation}\label{vh=ldr}
{L}_{\xi}f(t):=\lim_{\lambda \rightarrow
0}\frac{1}{\lambda}\{\phi_\lambda^* f(exp(\lambda
\xi)t)-f(t)\}=\frac{d}{d\lambda}|_{\lambda=0}(\phi_\lambda^*
f(t')),\quad t'=exp(\lambda \xi)t.
\end{equation}
Here $\phi_\lambda^*$ is the bull-back or the inverse differential
for the differential form or vector respectively.

Taking $\xi=\xi(t)=\xi_t$ in (\ref{vvft}), it follows that the Lie
derivative of $f(t)$ with respect to $\xi_t$ gives rise to the
horizontal variation of $f(t)$.

On the other hand, the time variation $\delta t$ can be expressed
by the contraction between the variational vector field
(\ref{vvft}) and 1-form $d_h t$ on $T^*{ T}$, i.e. $d t$ on $T^*{
E}$
\begin{equation}\label{vrvtime2}
i_{\xi_t} d_h t
=\delta t. 
\end{equation}

It is also feasible to express the  variation $\delta_h q^i(t)$ as
contraction of a horizontal variation vector field $\xi_h$ with
$d_v q^i$ or $d q^i$. To this purpose, $\xi_{h,q}$ along the fibre
with respect to horizontal variations of $q^i(t)$
\begin{equation}\label{vrvhq}
\xi_{h,q}:=\delta_h q^i(t) \frac{\partial}{\partial q^i}
\end{equation}
should be introduced. Combining with the vector field $\xi_t$ in
(\ref{vvft}), the general horizontal variational vector field
$\xi_h$ should be defined as
\begin{equation}\label{vrvh}
\xi_h:=\xi_t+\xi_{h,q}=\delta t\frac{\partial}{\partial
t}+\delta_h q^i(t) \frac{\partial}{\partial q^i}.
\end{equation}
Its contraction with $d_v q^i$ or $d q^i$ leads to
\begin{equation}\label{vrhq2}
i_{\xi_h}d q^i=d q^i \cdot \xi_h=\delta_h q^i(t). 
\end{equation}
In general, for any functional of $ q^i(t)$ and $\dot  q^i(t)$,
$F( q^i(t), \dot  q^i(t)): TQ \rightarrow R $, its (horizontal)
variation induced by (\ref{vrtnt}) is
\begin{equation}\begin{array}{rcl}\label{vrtf}
F( q^i(t), \dot  q^i(t))\rightarrow F( q^i(t'), \frac{d}{dt'}
q^i(t'))&=& F( q^i(t), \dot  q^i(t))+ \delta_h F( q^i(t),
\dot  q^i(t)),\\[2mm]
 \delta_h F( q^i(t), \dot
q^i(t))&=& i_{\xi_h} dF( q^i(t), \dot  q^i(t)).
\end{array}\end{equation}

\vskip 3mm {\it\bf Remark 2.3:}

For the total variation, a total variational vector field for
$q^i(t)$ along the fibre can also be introduced
\begin{eqnarray}\label{vrvvt}
\xi_{total}:=\xi_v+\xi_{h}&=&\delta t\frac{\partial}{\partial
t}+\delta_t q^i(t)\frac{\partial}{\partial q^i}\nonumber\\[2mm]
&=&\delta t\frac{\partial}{\partial t}+(\delta_v q^i(t)+\delta
t\frac{d}{dt}q^i(t))\frac{\partial}{\partial q^i},
\end{eqnarray}
whose contraction with $dq^i$ leads to the total variation
$\delta_t q^i(t)$
\begin{equation}\label{vrhq3}
i_{\xi_{total}}d q^i=dq^i \cdot \xi_{total}=\delta_t q^i(t).
\end{equation}

If we introduce the Lagrangian 1-from
\begin{equation}\label{Lform}
{\bf L}:=L(q^i, \dot q^i, t)dt
\end{equation}
and take $0=d^2 {\bf L}$, it is easy to see that the theorem 1
still holds. This means that  the total variations keep the
Euler-Lagrange cohomology as well as the necessary and sufficient
condition for symplectic structure preserving property in
classical mechanics.

\vskip 3mm {\it\bf Remark 2.4:}

In some literatures (see, for example, \cite{DR94}), it is
required that Hamilton's principle holds for the total variation
of the action, i.e. $\delta_t{ S}=0$, and regard $\delta_tq^i$ and
$\delta t$  as independent variations. Thus it follows the
Euler-Lagrange equation, the conservation relation for
the energy and the surface term %
\begin{equation}\label{eqclaw}\begin{array}{l}%
\frac{\partial {L}}{\partial q^i}-\frac{d}{dt}(\frac{\partial
{L}}{\partial \dot q^i})=0,\\
\frac{d}{dt} {H}+\frac{\partial}{\partial t}{L}=0,\\
\vartheta=\frac{\partial {L}}{\partial \dot q^j}\delta_t
q^j-{H}\delta t.
\end{array}
\end{equation}
If $\frac{\partial}{\partial t}{L}=0$, i.e. the system is
conservative, the energy ${H}$ is conserved. However,
$\delta_tq^i$ is actually dependent on $\delta t$. Therefore, it
would be better to regard $\delta_v q^i$ and $\delta t$ as
independent variations.

\subsection{Variations in Hamiltonian mechanics}

The action principle should, of course, be applied to the
Hamiltonian mechanics. In order to transfer to the Hamiltonian
formalism, we introduce a set of conjugate momenta from the
Lagrangian
${L}(q^i(t), \dot q^i(t);t)$ 
\begin{equation}\label{pm}
p_{j }= \frac {\partial {L}} {\partial {{\dot {q}}^j}},
\end{equation}
and take a Legendre transformation to get the Hamiltonian
\begin{equation}\label{lgdm}
{H}:={H}({q^i}, p_j; t )=p_k {{\dot q}^k} -{L}({q^i}, {\dot q}^j
;t).
\end{equation}

Now the action functional can be expressed as
\begin{equation}\label{smh1}
{S}([p_i(t)], [ q^i(t)]; t_1, t_2) =\int_{t_1}^{t_2}dt \{p_{k }
{\dot q}^k -{H}({q^i}, {p}_{j};t)\}.
\end{equation}
The total variation of the action can be calculated
\begin{equation}\label{vrsmh}
\begin{array}{rcl}
\delta_t{S}&=&\delta_v{S}+\delta_h{S},\\
\delta_v{S}&=& \int_{t_1}^{t_2}dt \{[{H}_{p_i}]%
\delta_v p_i  -[{H}_{q^i}]\delta_v q^i+\frac{d}{dt}(p_i
\delta_v q^i)\},\\
\delta_h{S}&=&\int_{t_1}^{t_2}dt \{[{H}_{p_i}]%
\delta_h p_i  -[{H}_{q^i}]\delta_h q^i+[ \frac{d}{dt} {H}
-\frac{\partial}{\partial t}{H}]\delta t+\frac{d}{dt}(p_i \delta_h
q^i-{H}\delta t)\},
\end{array}\end{equation}
where $[{H}_{p_i}]$, $[{H}_{q^i}]$ are canonical operators
\begin{equation}\label{cnlop}
[{H}_{p_i}]:=\dot q^i-\frac{\partial {H} } {\partial p_{i}},\qquad
[{H}_{q^i}]:=\dot p_{i}+\frac {\partial {H} } {\partial { q ^i}}.%
\end{equation}

 Thus, the stationary requirement for the vertical variation of the
 action $\delta_v{S}=0$ leads to the canonical equations
 \begin{equation}\label{vrsmh1}
\dot q^i=\frac{\partial {H} } {\partial p_{i}},\qquad \dot
p_{i}=-\frac {\partial {H} } {\partial { q ^i}}.
 \end{equation}
While the time re-parameterization invariance of the action, i.e.
$\delta_h{S}=0$ gives rise to an identity on the condition of the
energy
\begin{equation}\label{idh}
[{H}_{p_i}] \dot p_i  -[{H}_{q^i}]\dot q^i+ \frac{d}{dt} {H}
-\frac{\partial}{\partial t}{H}\equiv 0,
\end{equation}
and 
the boundary term that leads to so called ``extended symplectic
potential" is in fact a total divergence
\begin{equation}\label{exsym}
\int_{t_1}^{t_2}dt \frac{d}{dt}(p_i \delta_h q^i-{H}\delta
t)=\int_{t_1}^{t_2}dt \frac{d}{dt}({L}\delta t)=0.
\end{equation}

Similar to the Lagrangian mechanics, all remarks in last
subsection can be made for the Hamiltonian formalism. Especially,
the theorem 1 can also be established here.

\section{General Variations for Discrete Mechanics}

We have proposed a difference variational principle for the
(vertical) variation in discrete Lagrangian mechanics in view of
the differences with fixed time step-length being regarded as
entire variables \cite{glw}, \cite{glww}. One of advantages of
this approach is that it is available to both Lagrangian and
Hamiltonian formalism for the discrete mechanics. This approach
can also be generalized to the total variation for the differences
with variable time step-length being regarded as entire variables
in such a way that the variable time step-length should be
determined by an equation given by the variation problem with
variable discrete integral domain. In Lee's wards, discrete time
is regarded as a dynamical variable.

 Consider the case
that ``time" $t$ is difference discretized
\begin{equation}\label{td}
t\in R  \rightarrow t\in  {T_D}=\{ (t_k , t_{k+1}=t_k+\Delta t_k,
\quad k \in Z)\}
\end{equation}
and the step-lengths $\Delta t_k$ are determined by a variational
equation,  while the $n$-dimensional configuration space $M_k$ at
each moment $t_k, k \in Z$, is still continuous and smooth enough.

Let $N$ be the set of all nodes on ${T_D}$ with index set
$Ind({N})=Z$, ${M}=\bigcup_{k \in Z} M_k$ the configuration space
on $T_D$ that is 
at least pierce wisely
smooth enough. At the moment $t_k$, ${\cal N}_k$  be the set of
nodes neighboring to $t_k$.
 Let ${I}_k $ the index set of nodes of ${\cal N}_k$ including $t_k$.
  The coordinates of $M_k$
 are   denoted  by $q^i({t_k})=q^{i (k)}, i=1, \cdots,
n$.  $T(M_k)$ the tangent bundle of $M_k$ in the sense that
difference at $t_k$ is its base, $T^*(M_k)$ its dual. Let ${\cal
M}_k=\bigcup_{{l} \in {I}_k} M_l $ be the union of configuration
spaces $M_l$ at $t_l, {l} \in {I}_k$ on ${\cal N}_k$, $T{\cal
M}_k=\bigcup_{{l}\in {I}_k}TM_l $ the union of tangent bundles on
${M}_k$, $F(TM_k)$ and $F(T{\cal M}_k)$ the function spaces on
each of them respectively, etc.. In the difference variational
approach, we will use these notions.

\subsection{Variable difference Lagrangian mechanics}

$\quad$ Let us consider the system with a  discrete Lagrangian
${{L}_D}^{(k)}$  on $F(T({M}_k \times {T_D}))$.  For simplicity,
the Lagrangian is of the first order of differences
\begin{equation}\label{lmd}
{{L}_D}^{(k)}={L}_D( q^{i (k)}, {\Delta_{k} q}^{i (k)}; t_k),%
\end{equation}
with the difference ${\Delta_{k} q}^{i (k)}$ of $q^{i (k)}$ at $t_k$ defined by%
\begin{equation}\label{dfc}
{\Delta_{k} q}^{i (k)}:=\frac{q^{i (k+1)}-q^{i
(k)}}{t_{k+1}-t_{k}}.
\end{equation}
 The discrete action of the system is given by
\begin{equation}
{S}_D=\sum_{k \in Z} (t_{k+1}-t_k){{L}_D}^{(k)}(q^{i
(k)}, \frac{q^{i (k+1)}-q^{i (k)}}{t_{k+1}-t_k}; t_k).%
\end{equation}%

The discrete total variations for $q^{i (k)}=q^{i}(t_k)$
 should be defined as follows
\begin{equation}\label{tvdq}\begin{array}{rcl}
\delta_tq^{i (k)}&:=&q'^{i}(t'_k)-q^{i}(t_k)=\delta_v q^{i
(k)}+\delta_h q^{i (k)}, \quad
t'_k=t_k+\delta t_k,\\
\delta_v q^{i (k)}&:=&q'^i(t_k)-q^i (t_k),\qquad\qquad\qquad\qquad ~~\delta_v t_k=0,\\%
\delta_h q^{i (k)}&:=&q'^i(t'_k)-q'^i (t_k)=q^i(t'_k)-q^i
(t_k)+O(\delta^2).
\end{array}\end{equation}
It can be shown that horizontal variation $\delta_h q^{i (k)}$ is
given by
\begin{equation}\label{hvdq}
\delta_h q^{i (k)}=\delta t_k \Delta_k q^{i (k)}.%
\end{equation}
The discrete total variations for ${\Delta_k q}^{i (k)}$ are
defined as
\begin{equation}\label{tvDq}\begin{array}{l}
\delta_t\Delta_k q^{i
(k)}:=\frac{q'^i(t'_{k+1})-q'^i(t'_{k})}{t'_{k+1}-t'_k}-\frac{q^i(t_{k+1})-q^i(t_{k})}{t_{k+1}-t_k}=\delta_v
\Delta_k q^{i (k)}+\delta_h
\Delta_k q^{i (k)},\\%
\delta_v \Delta_k q^{i (k)}:=\frac{q'^i(t_{k+1})-q'^i(t_{k})}{t_{k+1}-t_k}-\frac{q^i(t_{k+1})-q^i(t_{k})}{t_{k+1}-t_k}.
\end{array}\end{equation}
Due to the definition of the difference with variable time
step-length (\ref{dfc}) and the Leibniz law for it
\begin{equation}\label{lbnz}
\Delta_k (f^{(k)}g^{(k)})=(\Delta_k f^{(k)})g^{(k)}+
Ef^{(k)}(\Delta_k g^{(k)}),
\end{equation}
where $E$ is the shift operator defined as
\begin{equation}\label{shft}
Ef^{(k)}=f^{(k+1)},\qquad E^{-1}f^{(k)}=f^{(k-1)},
\end{equation}
 it follows
\begin{equation}\label{tvDq2}\begin{array}{l}
\delta_t\Delta_k q^{i (k)}=\Delta_k (\delta_tq^{i (k)})-(\Delta_k
\delta t_k) \Delta_k q^{i (k)},\\%
\delta_h \Delta_k q^{i (k)}=\delta t_{k+1}\Delta(\Delta_k q^{i (k)}).%
\end{array}\end{equation}
Namely,
\begin{equation}\label{tvDq3}\begin{array}{l}
\delta_v \Delta_k q^{i (k)}=\Delta_k \delta_v q^{i (k)},\\
\delta_h \Delta_k q^{i (k)}=\Delta_k \delta_h q^{i (k)} -(\Delta_k
\delta t_k) \Delta_k q^{i (k)}.%
\end{array}\end{equation}

Using above properties and
\begin{equation}\label{dDt}
\delta_t(t_{k+1}-t_k)=\Delta_k(\delta_tt_k)(t_{k+1}-t_k),
\end{equation}
the total variations of the discrete Lagrangian can be calculated
as follows
\begin{equation}\label{tvdL}\begin{array}{rcl}
\delta_t{{L}_D}^{(k)}&=&\frac{\partial{{ L}_D}^{(k)}}{\partial
q^{i (k)}}\delta_tq^{i (k)}+\frac{\partial{{L}_D}^{(k)}}{\partial
\Delta_k q^{i (k)}}\delta_t\Delta_k q^{i (k)}+\frac{\partial{{
L}_D}^{(k)}}{\partial t_k} \delta_t t_k \\
&=&[L_{q^{i (k) }}]\delta_tq^{i (k)}\\
&+&\Delta_k({p_i}^{(k)} \Delta_k  q^{i (k-1)})\delta
t_k+\frac{\partial{{
L}_D}^{(k)}}{\partial t_k} \delta t_k \\
&+&\Delta_k ( {p_i}^{(k+1)} \delta_tq^{i (k)}-{p_i}^{(k)} \Delta_k
q^{i (k-1)}\delta t_k),
\end{array}\end{equation}
where $[L_{q^{i (k) }}]$ is the discrete Euler-Lagrange operator
\begin{equation}\label{eloD}
[L_{q^{i (k) }}]:=\frac{\partial{{ L}_D}^{(k)}}{\partial q^{i
(k)}}-\Delta ( \frac{\partial{{ L}_D}^{(k-1)}}{\partial \Delta
q^{i (k-1)}}),
\end{equation}
and ${p_i}^{(k)}$ the  discrete canonical conjugate momenta
\begin{equation}\label{mmntad}
{p_i}^{(k)}:=\frac{\partial{{L}_D}^{(k-1)}}{\partial \Delta q^{i
(k-1)}}.
\end{equation}

Thus the total variation of  action is given by
\begin{equation}\label{tvdS}\begin{array}{rcl}
\delta_tS_D&=&\sum_k (t_{k+1}-t_k)\{(\Delta \delta t_k){{
L}_D}^{(k)}+\delta_t{{L}_D}^{(k)}\}\\
&=&\sum_k (t_{k+1}-t_k)\{[L_{q^{i (k) }}]\delta_tq^{i (k)}
+(\Delta_k{H_D}^{(k-1)} +\frac{\partial{{
L}_D}^{(k)}}{\partial t_k}) \delta t_k \\
&+&\Delta_k ({p_{i}}^{(k+1)}
\delta_tq^{i (k)}-{H_D}^{(k-1)} \delta t_k)\},
\end{array}\end{equation}
where ${{H}_D}^{(k)}$ is the difference Hamiltonian that can be
introduced through the discrete Legendre transformation
\begin{equation}\label{lgdrd}
{{H}_D}^{(k)}:={p_{i}}^{(k+1)}\Delta_t q^{i
(k)}-{{L}_D}^{(k)}.%
\end{equation}
Thus the total variation of the discrete action (\ref{tvdS}) can
be written as
\begin{equation}\label{tvds2}\begin{array}{rcl}
\delta_t{S}_D&=&\delta_v {S}_D+\delta_h {S}_D,\\
\delta_v{S}_D&=&\sum_k (t_{k+1}-t_k)[L_{q^{i (k) }}]\delta_v q^{i
(k)}+\Delta ({p_{i}}^{(k+1)}
\delta_vq^{i (k)}),\\
\delta_h {S}_D&=&\sum_k (t_{k+1}-t_k)\{[L_{q^{i (k) }}]\delta_h
q^{i (k)}+(\Delta{{H}_D}^{(k-1)} +\frac{\partial{{
L}_D}^{(k)}}{\partial t_k}) \delta t_k \\&+&\Delta
({p_{i}}^{(k+1)} \delta_h q^{i (k)}- {{H}_D}^{(k-1)}\delta t_k)\}.
\end{array}\end{equation}

The variational principle requires $\delta_v {S}_D=0$ and the
discretized re-parameterization invariance with respect to
discrete time may also leads to $\delta_h{S}_D=0$ if this
invariance does exist. Thus it follows the discrete Euler-Lagrange
equations for $q^{i (k)}$'s
\begin{equation}\label{eleqd}
\frac{\partial{{L}_D}^{(k)}}{\partial q^{i (k)}}-\Delta (
\frac{\partial{{L}_D}^{(k-1)}}{\partial \Delta q^{i
(k-1)}})=0,%
\end{equation}
and the equation for the variable time step-length
\begin{equation}\label{timestep}
(\frac{\partial{{L}_D}^{(k)}}{\partial q^{i (k)}}-\Delta (
\frac{\partial{{L}_D}^{(k-1)}}{\partial \Delta q^{i
(k-1)}}))\Delta q^{i (k)}+\Delta {{H}_D}^{(k-1)}
-\frac{\partial{H_D}^{(k)}}{\partial t_k}=0.%
\end{equation}

It is more or less straightforward to show that if the time
step-length is fixed the equation (\ref{timestep}) has no solution
in general even if the Lagrangian does not depend on discrete time
manifestly. In other wards, for the conservative discrete
Lagrangian mechanics the time step-length should be variable in
general so as to the energy of the system can be kept conserved
discretely.
\vskip 3mm {\bf Remark 3.1:}

We may introduce exterior differential operators $\hat{d}$, $d_v$
and $\hat{d}_h$ on $T^*(M \times T_D)$, $T^*M$ and $T^* T_D$
respectively. They are nilpotent and satisfy
\begin{equation}\label{dhat}
\hat{d}=d_v+\hat{d}_h,\qquad \{d_v, \hat{d}_h\}=0.
\end{equation}
Especially, $\hat{d}_h$ is due to the difference on $T_D$ and
satisfy  Leibniz's law for ordinary forms {\footnote{It is needed
some noncommutative differential calculus to completely clarify
the properties of $\hat{d}_h$. For the case that $\Delta t$ is
fixed, the noncommutative differential calculus can be found in
\cite{gwwww},\cite{gwz}. For the case of variable time steps,
similar noncommutative differential calculus can be established
\cite{gw}.}}.

\vskip 3mm {\bf Remark 3.2:}

Actually, analog to the case with fixed time steps \cite{glw},
\cite{glww}, it can be established the difference version for the
Euler-Lagrange cohomology and the necessary and sufficient
condition for the difference conservation law of the discrete
symplectic 2-form.

From $\delta_v S_D$ in (\ref{tvds2}), it is easy to see that we
may take $d_v$ on $S_D$ to get
\begin{equation}\label{dSD}
d_v S_D=\sum_k (t_{k+1}-t_k)d_v {L_D}^{(k)}, \qquad d_v
{L_D}^{(k)}={{\cal E}_D}^{(k)}+\Delta_k
{\theta_D}^{(k)},
\end{equation}
where ${{\cal E}_D}^{(k)}$, ${\theta_D}^{(k)}$ are the discrete
Euler-Lagrange 1-form and symplectic potential 1-form respectively
\begin{eqnarray}
{{\cal E}_D}^{(k)}:=[L_{q^{i (k) }}]d_v q^{i (k)},\qquad
{\theta_D}^{(k)}:={p_{i}}^{(k+1)} d_vq^{i (k)}.%
\end{eqnarray}
Then due to the nilpotency of $d_v$, it is straightforward to get
\begin{equation}\label{ddsD}
d_v{{\cal E}_D}^{(k)}+\Delta_k {\omega_D}^{(k)}=0,\qquad
{\omega_D}^{(k)}:=d_v{\theta_D}^{(k)}=d_v{p_{i}}^{(k+1)}\wedge
d_vq^{i
(k)}.
\end{equation}
Therefore, we may get the discrete version for the theorem 1
\cite{glw}, \cite{glww1}, \cite{glww}:

\vskip 3mm{\it Theorem 2: For all discrete Lagrangian of a kind of
 systems with first order differences on the bundle $E(T_D, Q, \pi) \simeq M
\times T_D$,

1. The following discrete version of the Euler-Lagrange cohomology
is nontrivial: {\centerline{$H_{DCM}$:=\{Closed Euler-Lagrange
forms\}/ \{Exact Euler-Lagrange forms\}.}}

 2. The necessary and
sufficient condition for conservation of the discrete symplectic
2-form, i.e.
\begin{equation}\label{csvsyD}
\Delta_k{\omega_D}^{(k)}=0,
\end{equation}
is the corresponding discrete Euler-Lagrange 1-form being closed.
}

\vskip 3mm {\bf Remark 3.3:}

In this paper, $T_D$ is an infinite chain. It is reasonable to
consider an interval on $T_D$. We will publish the issues on this
topic elsewhere.

\subsection{Variable difference Hamiltonian mechanics}

Now we consider the total difference variation on the phase space
in the  discrete  Hamiltonian formalism with variable (time
step-length) difference.

In order to transfer to the discrete  Hamiltonian formalism, it is
needed to introduce the discrete canonical conjugate momenta
according to the equation (\ref{mmntad}) and express the discrete
Lagrangian by the discrete Hamiltonian via  Legentre
transformation (\ref{lgdrd}). Thus, the discrete action can be
expressed as
\begin{equation}\label{actndh}\begin{array}{rcl}
{S}_D&=&\sum_{k}(t_{k+1}-t_k){{L}_D}^{(k)}(q^{i (k)},
\frac{q^{i (k+1)}-q^{i (k)}}{t_{k+1}-t_k}, t_k)\\
&=&\sum_{k}(t_{k+1}-t_k)({p_{i}}^{(k+1)}\Delta_t q^{i (k)}-{{
H}_D}^{(k)}).
\end{array}\end{equation}
And its total variation reads
\begin{equation}\label{tvdsh}\begin{array}{rcl}
\delta_t{S}_D&=&\delta_v {S}_D+\delta_h {S}_D\\
&=&\sum_k (t_{k+1}-t_k)\{(\Delta q^{i (k)}- \frac{\partial{{\cal
H}_D}^{(k)}}{\partial {p_i}^{(k+1)}})\delta_t{p_i}^{(k+1)}\\
&&
-(\Delta {p_i}^{(k)}+\frac{\partial{{H}_D}^{(k)}}{\partial {q}^{i
(k)}})\delta_t q^{i (k)}
+(\Delta{{H}_D}^{(k-1)} +\frac{\partial{{
L}_D}^{(k)}}{\partial t_k}) \delta t_k \\
&&+\Delta ({p_i}^{(k)} \delta_tq^{i (k)}- {{ H}_D}^{(k-1)}\delta
t_k)\}.
\end{array}\end{equation}

Similar to the discrete Lagrangian formalism, Hamilton's 
principle requires $\delta_v {S}_D=0$ and the discretized
re-parameterization invariance with respect to discrete time may
also lead to $\delta_h{S}_D=0$ if this invariance does exist. Thus
it follows the discrete canonical equations for ${p_i}^{(k)}$'s
and $q^{i (k)}$'s
\begin{equation}\label{cneqd}
\Delta q^{i (k)}= \frac{\partial{{H}_D}^{(k)}}{\partial
{p_i}^{(k+1)}},\quad \Delta {p_i}^{(k)}=-\frac{\partial{{
H}_D}^{(k)}}{\partial {q}^{i (k)}},%
\end{equation}
 the equation for the variable time step-length
\begin{equation}\label{timesteph}\begin{array}{rcl}
(\Delta q^{i (k)}- \frac{\partial{{H}_D}^{(k)}}{\partial
{p_i}^{(k+1)}})\Delta {p_i}^{(k+1)} -(\Delta
{p_i}^{(k)}+\frac{\partial{{H}_D}^{(k)}}{\partial {q}^{i
(k)}})\Delta q^{i (k)}+\Delta {{H}_D}^{(k-1)}
-\frac{\partial{{H}_D}^{(k)}}{\partial t_k}=0.%
\end{array}\end{equation}

 It is also more or less straightforward to show that if
the time step-length is fixed the equation (\ref{timesteph}) has
no solution in general even if the Hamiltonian does not depend on
discrete time manifestly. In other wards, for the conservative
discrete Hamiltonian mechanics the time step-length should be
variable so as to the energy of the system can be kept conserved
discretely. In \cite{td1}, \cite{td2}, \cite{td3}, \cite{gm88}, \cite{kmo99},
 this issue has been studied.

It should be mentioned that all remarks in last subsection can be
made here and the theorem 2 can also be established for the
discrete Hamiltonian formalism.

\section{General Variations for Field Theory}

Consider a bundle $E(X, Q, \pi)$, the fibre $Q\simeq M$. For simplicity, let 
$X=X^{(1,n-1)}$ be an $n$-dimensional Minkowskian space as base
manifold with coordinates $x^{\mu}$, $(\mu =0, \cdots,n-1 )$, $M$
the configuration space
 on $X^{(1,n-1)}$ with a set of generic (scalar) fields $u^{\alpha}(x)$,
 $(\alpha=1, \cdots, r)$,
$TM$ the tangent bundle of $M$ with coordinates $(u^{\alpha},
u_{\mu}^{\alpha})$, where
 $u_{\mu}^{\alpha}=\frac {\partial u^{\alpha}} {\partial x^{\mu}}$,
 $F(TM)$ the function space on $TM$ etc.

 We also assume
these fields to be  free of constraints. In fact, the approach
here can easily be applied to other cases.

\subsection{General variation in Lagrangian formalism}

The Lagrangian of the theory is supposed to be the first order of
derivatives of the fields and dependent to the coordinates
manifestly, i.e.
${L}(u^{\alpha}, u_{\mu}^{\alpha}; x^\mu )$, 
and the action is
\begin{equation}\label{actnft}
{S}([u^\alpha(x)]; x^\mu)=\int_\Omega d^4 x {L}(u^{\alpha},
u_{\mu}^{\alpha}; x^\mu ).
\end{equation}

Let us consider the variations of the fields, i.e. total variation
$\delta_tu^{\alpha}$, vertical one $\delta_v u^{\alpha}$ and
horizontal one $\delta_h u^{\alpha}$:
\begin{equation}\begin{array}{l}\label{vru}
u^{\alpha}\rightarrow u'^{\alpha}(x')=u^{\alpha}(x)+\delta_t
u^{\alpha}(x),\\
\qquad \delta_tu^{\alpha}=\delta_v u^{\alpha}+\delta_h u^{\alpha},\\
\qquad \delta_v u^{\alpha}(x):= u'^{\alpha}(x)-u^{\alpha}(x),\\
\qquad \delta_h
u^{\alpha}(x):=u'^{\alpha}(x')-u'^{\alpha}(x)=u^{\alpha}(x')-u^{\alpha}(x)+O(\delta^2)\\
\qquad\qquad\qquad=\delta x^\mu
\partial_\mu u^{\alpha}(x),
 \end{array}\end{equation}
 accompanying with the coordinates' infinitesimal continuous
transformation
\begin{equation}\label{vrx}
x^\mu\rightarrow x'^\mu=x^\mu +\delta x^\mu.
\end{equation}
The corresponding changes in the derivative of fields
$u_\mu^{\alpha}$ are
\begin{equation}\begin{array}{l}\label{vrpau}
\frac{\partial}{\partial x^\mu}u^{\alpha}(x)\rightarrow
\frac{\partial}{\partial
x'^\mu}u'^{\alpha}(x')=\frac{\partial}{\partial
x^\mu}u^{\alpha}(x)+\delta_t(\frac{\partial}{\partial
x^\mu}u^{\alpha}(x)),\\
\qquad \delta_t(\frac{\partial}{\partial
x^\mu}u^{\alpha}(x))=\delta_v(\frac{\partial}{\partial
x^\mu}u^{\alpha}(x))+\delta_h(\frac{\partial}{\partial
x^\mu}u^{\alpha}(x)),\\
\qquad \delta_v(\frac{\partial}{\partial x^\mu}u^{\alpha}(x)):=
\frac{\partial}{\partial x^\mu}u'^{\alpha}(x)-\frac{\partial}{\partial x^\mu}u^{\alpha}(x),\\
\qquad \delta_h(\frac{\partial}{\partial x^\mu}u^{\alpha}(x))
=\delta x^\nu \frac{\partial}{\partial x^\nu} (\frac{\partial}{\partial
x^\mu}u^{\alpha}(x)).
\end{array}\end{equation}

Now the action (\ref{actnft}) is also changed as follows
\begin{equation}\begin{array}{l}\label{vractn}
{S}([u^\alpha(x)]; x^\mu)\rightarrow { S'}([{u'}^{\alpha}(x')];
x'^\mu)=\int_{\Omega'} d^n x' {L}'({u'}^{\alpha}(x'), {u'}_{{\mu}'}^{\alpha}(x'); x'^\mu )\\
\qquad\qquad\qquad\qquad=\int_{\Omega} d^n x \det(\frac{\partial
x'}{\partial x}) \{{L}({u}^{\alpha}(x),
{u}_{{\mu}}^{\alpha}(x); x^\mu )+\delta_t{L}\}\\
\qquad\qquad\qquad\qquad ={S}([u^\alpha(x)];
x^\mu)+\delta_t{S}. 
\end{array}\end{equation}
Using Jacobi formula for the measure
\begin{equation}\label{jcb}
d^nx'=\det (\frac{\partial x'}{\partial x}) d^n
x=(1+\frac{\partial \delta x^\mu}{\partial x^\mu})d^n x,%
\end{equation}
we get
\begin{equation}\begin{array}{rcl}\label{vrsfl}
\delta_t{S}([u^\alpha(x)]; x^\mu)&=&\int_\Omega d^nx
\{\partial_\mu \delta x^\mu{L}+\delta_t{L}\}\\
&=&\int_\Omega d^nx\{[{L}_{u^\alpha}]\delta_t
u^\alpha+(\partial^\mu
T_{\mu\nu}+\frac{\partial {L}}{\partial x^\nu })\delta x^\nu\\
&& +\partial_\mu(\frac{\partial {L}}{\partial(\partial_\mu
u^\alpha)}\delta_tu^\alpha-{T^\mu}_\nu \delta x^\nu)\},
\end{array}\end{equation}
\begin{equation}\begin{array}{rcl}\label{vrsfv}
\delta_v{S}([u^\alpha(x)]; x^\mu)&=&\int_\Omega d^nx\{[{
L}_{u^\alpha}]\delta_v u^\alpha+\partial_\mu(\frac{\partial {
L}}{\partial(\partial_\mu u^\alpha)}\delta_v u^\alpha)\},
\end{array}\end{equation}
\begin{equation}\begin{array}{rcl}\label{vrsfh}
\delta_h{S}([u^\alpha(x)]; x^\mu)&=&\int_\Omega d^nx\{[{
L}_{u^\alpha}]\delta_h u^\alpha+(\partial^\mu
T_{\mu\nu}+\frac{\partial {L}}{\partial x^\nu })\delta x^\nu\\
&& +\partial_\mu(\frac{\partial {L}}{\partial(\partial_\mu
u^\alpha)}\delta_h u^\alpha-{T^\mu}_\nu \delta x^\nu)\}\\
&=&\int_\Omega d^nx \partial_\mu({L}\delta x^\mu),
\end{array}\end{equation}
where $[{L}_{u^\alpha}]$, $T_{\mu\nu}$ are the Euler-Lagrange
operator and energy-momentum tensor respectively
\begin{equation}\begin{array}{l}\label{dfn}
[{L}_{u^\alpha}]:=\frac{\partial {L}}{\partial
u^\alpha}-\partial_\mu(\frac{\partial {L}}{\partial
u_\mu^\alpha}),\\
\quad T_{\mu\nu}:=\frac{\partial {L}}{\partial (\partial^\mu
u^\alpha)}\partial_\nu u^\alpha -{L}\eta_{\mu\nu}.
\end{array}\end{equation}
Thus $\delta_t{S}=0$, i.e. $\delta_v{S}=0$ due to Hamilton's
principle and $\delta_h{ S}=0$ due to the invariance of
re-parameterization of the coordinates that suppose to keep the
metric {\footnote{Thus the symmetry is the Poincar$\grave{e}$
transformations. In general, the general coordinate
transformations can be considered and all formulae become
covariant.}}, requires to regard $\delta_v u^\alpha$ and $\delta
x^\mu$ as independent components and gives rise to the
Euler-Lagrange equation
\begin{equation}\label{eleq}
\frac{\partial {L}}{\partial u^\alpha}-\partial_\mu(\frac{\partial
{L}}{\partial
u_\mu^\alpha})=0%
\end{equation}
as well as an identity between the Euler-Lagrange operator and
conservation property for the energy-momentum tensor
\begin{equation}\label{emcnsv}
(\frac{\partial {L}}{\partial
u^\alpha}-\partial_\mu(\frac{\partial {L}}{\partial
u_\mu^\alpha}))\partial_\nu u^\alpha+\partial^\mu
T_{\mu\nu}+\partial_\nu{L}=0.
\end{equation}
This equation shows that the energy-momentum tensor is conserved
if and only if the Euler-Lagrange equation is satisfied and the
Lagrangian does not depend manifestly on the coordinates.

The boundary term is as follows
\begin{equation}\begin{array}{l}\label{bdry}
\int_{\Omega}d^nx \frac{\partial}{\partial x^\mu}((\frac{\partial
{L}}{\partial u_\mu^\alpha})\delta_v u^\alpha+{L}\delta
x^\mu)=\int_{\partial\Omega}((\frac{\partial {L}}{\partial
u_\mu^\alpha})\delta_v u^\alpha+{L}\delta
x^\mu)d\sigma^\mu=0.%
\end{array}\end{equation}%
 If we define a current
\begin{equation}\label{crt}
{J}^\mu:=(\frac{\partial {L}}{\partial
u_\mu^\alpha})\delta_v u^\alpha+{L}\delta x^\mu,%
\end{equation}
the equation (\ref{bdry}) leads to an continuity equation for the
current if and only if the Euler-Lagrange equation is satisfied:
\begin{equation}\label{crtcsv}
\frac{\partial}{\partial x^\mu}{J}^\mu=0.
\end{equation}
In  fact, this current is just the Noether current with respect to
the invariance of the action under re-parameterization of the
right-angle coordinates on $M$.

\vskip 3mm {\bf Remark 4.1:}

Introducing an exterior differential operator
 $d_v$ along the fibre $Q \simeq M$ that
satisfies
\begin{equation}\label{dv1}
{d_v}^2=0, \qquad \{d_v, d_h\}=0, \qquad d:=d_v+d_h,
\end{equation}
where $d_h$ and $d$ is the nilpotent exterior differential
operator on $T^*{X}$ and $T^*E$, $E=M \times X$ respectively as
was as
a vertical variational vector field
\begin{equation}\label{vvvfu}
\xi_u:=\delta_v u^\alpha (x) \frac{\partial}{\partial u^\alpha},
\end{equation}
then
\begin{equation}\label{vvu}
\delta_v u^\alpha(x) =i_{\xi_u}d_v u^\alpha=i_{\xi_u}du^\alpha.
\end{equation}

By means of the vertical variational vector field (\ref{vvvfu}) on
$TQ$, $\delta_v{S}$ can also be expressed as its contraction with
1-form $d_v{S} \in T^* Q $ %
\begin{equation}\label{vrs1}
i_{\xi_u} d_v{ S}=\delta_v{S}. %
\end{equation}
We may calculate $d_v{ S} \in T^* Q$. Since $d_v$ commutes with
the integral and $d^n x$ (see also, for example, the functional
differential calculus in \cite{olv93}), it is straightforward to
get
\begin{equation}\label{dvsf1}
 d_v{ S}=\int_\Omega d^n x \{[{ L}_{u^\alpha}] d_v u^\alpha
+\frac{\partial}{\partial x^\mu}\theta^\mu
)\},%
\end{equation}
where $\theta^\mu$ are the Lagrange
1-forms%
\begin{equation}
\theta^\mu:=\frac{\partial { L}}{\partial u^\alpha_\mu}d_v u^\alpha.%
\end{equation}
Now by contracting with the vertical variational vector field
(\ref{vvvfu}) it follows straightforwardly  $\delta_v S$.

Furthermore, due to the nilpotency of $d_v$, it is easy to
get\begin{equation}\label{dEu} d_v{\cal
E}_u+\frac{\partial}{\partial x^\mu}\omega^\mu=0,
\end{equation}
where ${\cal E}_u$ is the Euler-Lagrange 1-form defined by
\begin{equation}\label{elff}
{\cal E}_u(u^\alpha(x), u^\alpha_\mu(x); x):=[{ L}_{u^\alpha}] d_v
u^\alpha,
\end{equation}
$\omega^\mu$ are the multi-symplectic 2-forms and in local
coordinates:
\begin{equation}\label{defspf}
\omega^\mu:=d_v \theta^\mu=\frac {\partial^2 { L}} {{\partial
u^\alpha}
\partial {u^\beta_\mu} } du^\alpha \wedge du^\beta
+\frac {\partial^2 { L}} {\partial {u^\alpha_\nu}{\partial
{u^\beta_\mu}}} du^\alpha_\nu \wedge d u^\beta.
\end{equation}
From the definition (\ref{elff}), equations (\ref{dvsf1}) and
(\ref{dEu}), the following theorem \cite{glw}, \cite{glww1},
\cite{glww} holds:

\vskip 3mm {\it Theorem 3: For all Lagrangian of a kind of field
theories with first order of derivatives on the bundle $E(X, Q,
\pi)$, 

1. The following Euler-Lagrange cohomology is  nontrivial:
$$H_{CFT}:=\{{\cal E}_u |d{\cal E}_u=0\}/ \{{\cal E}_u|{\cal E}_u=d\beta\},$$
where $\beta=\beta(u^\alpha(x), u^\alpha_\mu(x); x)$ is an
arbitrary function of $(u^\alpha(x), u^\alpha_\mu(x); x)$.

2. The necessary and sufficient condition for conservation of the
multi-symplectic 2-forms, i.e.
\begin{equation}\label{csvmsy}
\frac{\partial}{\partial x^\mu}\omega^\mu=0,
\end{equation}
is the corresponding Euler-Lagrange 1-form being closed. }

\vskip 3mm
 {\it\bf Remark 4.2:}

  From the definition of the Lie
derivative it can be seen that the horizontal variations are given
by the Lie derivative with respect to the variational vector
field.

Let $\xi$ be a vector field on $X$, $exp(\lambda \xi)$ be the flow
with parameter $\lambda$, i.e. the one-parameter diffeomorphism
group, induced by $\xi$, $f$ a differential or a vector on $X$.
The infinitesimal change of $f$ under flow is described by its Lie
derivative  with respect to the vector field $\xi$
\begin{equation}\label{vhf=ldr}
{L}_{\xi}f(x):=\lim_{\lambda \rightarrow
0}\frac{1}{\lambda}\{\phi_\lambda^* f(exp(\lambda
\xi)x)-f(x)\}=\frac{d}{d\lambda}|_{\lambda=0}(\phi_\lambda^*
f(x')),\quad x'=exp(\lambda \xi)x.
\end{equation}
Here $\phi_\lambda^*$ is the bull-back or the inverse differential
for the differential form or vector respectively.

Taking horizontal variational vector field for the coordinates
$\xi=\xi(x)=\xi_x$
\begin{equation}\label{vvfx}
\xi(x):=\delta x^\mu \frac{\partial}{\partial x^\mu}, 
\end{equation}
it follows that the Lie derivative of $f(x)$ with respect to
$\xi_x$ gives rise to the horizontal variation of $f(x)$.

On the other hand, the coordinate variations $\delta x^\mu$ can be
expressed by the contraction between the variational vector field
(\ref{vvfx}) and 1-form $d_h x^\mu$ on $T^*{ X}$, i.e. $d x^\mu$
on $T^*{ E}$
\begin{equation}\label{vrvx2}
i_{\xi_x} d_h x^\mu
=\delta x^\mu. 
\end{equation}
It is also feasible to express the  variation $\delta_h
u^\alpha(x)$ as contraction of a horizontal variation vector field
$\xi_h$ with $d_v u^\alpha$ or $d u^\alpha$. To this purpose,
$\xi_{h,u}$ along the fibre with respect to horizontal variations
of $u^\alpha(x)$
\begin{equation}\label{vrvhu}
\xi_{h,u}:=\delta_h u^\alpha(x) \frac{\partial}{\partial u^\alpha}
\end{equation}
should be introduced. Combining with the vector field $\xi_x$ in
(\ref{vvfx}), the general horizontal variational vector field
$\xi_h$ should be defined as
\begin{equation}\label{vrvhf}
\xi_h:=\xi_x+\xi_{h,u}=\delta x^\mu\frac{\partial}{\partial
x^\mu}+\delta_h u^\alpha(x) \frac{\partial}{\partial u^\alpha}.
\end{equation}
Its contraction with $d_v u^\alpha$ or $d u^\alpha$ leads to
\begin{equation}\label{vrhu2}
i_{\xi_h}d u^\alpha=d u^\alpha \cdot \xi_h=\delta_h u^\alpha(t). 
\end{equation}
In general, for any functional of $ u^\alpha(x)$ and
$u^\alpha_\mu(x)$, $F( u^\alpha(x), u^\alpha_\mu(x)): TQ
\rightarrow R $, its (horizontal) variation induced by (\ref{vrx})
is
\begin{equation}\begin{array}{rcl}\label{vrtff}
F( u^\alpha(x), u^\alpha(x))\rightarrow F( u^\alpha(x'),
\frac{\partial}{\partial x'^\mu} u^\alpha(x'))&=& F( u^\alpha(x),
u^\alpha_\mu(x))+ \delta_h F( u^\alpha(x),
u^\alpha_\mu(x)),\\[2mm]
 \delta_h F( u^\alpha(x), u^\alpha_\mu(x))&=& i_{\xi_h} dF( u^\alpha(x),
 u^\alpha_\mu(x)). %
\end{array}\end{equation}

\vskip 3mm {\it\bf Remark 4.3:}

For the total variation, a total variational vector field for
$u^\alpha(x)$ along the fibre can also be introduced
\begin{eqnarray}\label{vrvvtt}
\xi_{total}:=\xi_v+\xi_{h}&=&\delta x^\mu\frac{\partial}{\partial
x^\mu}+\delta_t u^\alpha(x)\frac{\partial}{\partial u^\alpha}\nonumber\\[2mm]
&=&\delta x^\mu\frac{\partial}{\partial x^\mu}+(\delta_v
u^\alpha(x)+\delta x^\mu\frac{\partial}{\partial x^\mu
}u^\alpha(x))\frac{\partial}{\partial u^\alpha},
\end{eqnarray}
whose contraction with $du^\alpha$ leads to the total variation
$\delta_t u^\alpha(x)$
\begin{equation}\label{vrhu3}
i_{\xi_{total}}d u^\alpha=du^\alpha \cdot \xi_{total}=\delta_t
u^\alpha(x).
\end{equation}



\subsection{General variation in Hamiltonian formalism}

In order to transfer to the Hamiltonian formalism for
classical field theory, we first define  a set of ``momenta"
canonically conjugate to
the field variables%
\begin{equation} \pi_{\alpha}(x)=\frac {\partial {L}} {\partial \dot {u}^{\alpha}},
\end{equation}
 and take a Legendre
transformation to get the Hamiltonian density
\begin{equation}\label{ltfmft}
{H}(u^{\alpha}, \pi_{\alpha }, {\nabla_a u}^{\alpha})=\pi_{\alpha
}(x) {\dot u}^{\alpha}(x) - {L}(u^{\alpha}, {\dot u}^{\alpha},
{\nabla_a u}^{\alpha}),
\end{equation}
where $\nabla_a =\frac{\partial}{\partial x^a}, a=1, \cdots, n-1$.
The Hamiltonian then is given by
\begin{equation}
 H(t)=\int_\Sigma d^{n-1}x
{H}(x), \end{equation}
 with the Legendre
transformation
\begin{equation}
 H(t)=\int_\Sigma d^{n-1}x \{\pi_{\alpha}(x)
{\dot u}^{\alpha}(x) -L(t)\},\qquad L(t)=\int_\Sigma d^{n-1}x {L},
\end{equation}
where $\Sigma\subset \Omega$ is a 3-dimensional simultaneous space-like
hypersurface in $\Omega$.

The action ${S}([u^\alpha(x)]; x^\mu)$ (\ref{actnft}) becomes 
\begin{equation}
{S}([u^\alpha(x)]; x^\mu)=\int_\Omega d^n x\{\pi_{\alpha}(x) {\dot
u}^{\alpha}(x) - {H}(u^{\alpha}, {\dot u}^{\alpha},{\nabla_a
u}^{\alpha})\}.
\end{equation}
The total variation of the action can be calculated similar to
that in the last subsection, but here $\pi_{\alpha}(x)$,
$u^\alpha(x)$ and their derivatives should be varied
independently. Then we get
\begin{equation}\begin{array}{rcl}\label{vrsf2}
\delta_t{S}([u^\alpha(x)]; x^\mu)&=&\int_\Omega d^n x
\{\partial_\mu \delta x^\mu{L}+\delta_t{L}\}\\[2mm]
&=&\int_\Omega d^n x\{[{H}_{\pi^\alpha}]\delta_t
\pi^\alpha-[{H}_{u^\alpha}]\delta_tu^\alpha\\[2mm]
&&+(\partial^\mu T_{\mu\nu}+\frac{\partial {H}}{\partial x^\nu
})\delta x^\nu -\partial_\mu(\frac{\partial {
H}}{\partial(\partial_\mu u^\alpha)}\delta_tu^\alpha-{T^\mu}_\nu
\delta x^\nu)\},
\end{array}\end{equation}
where $[{H}_{u^\alpha}]$, $[{H}_{\pi^\alpha}]$ are the canonical
operators
\begin{equation}\begin{array}{l}\label{cnlopf}
[{H}_{\pi_\alpha}]:={\dot u}^{\alpha}(x)-
 \frac{\partial {H}}{\partial \pi_{\alpha
}},\\[2mm]
[{H }_{u^\alpha}]:={\dot \pi}_{\alpha}(x)+\frac{\partial {
H}}{\partial u^{\alpha}}-\nabla_a (\frac {\partial{
H}}{\partial(\nabla_a u^{\alpha})}).
\end{array}\end{equation}

Similar to the case of Lagrangian formalism, $\delta_t{S}=0$, i.e.
$\delta_v{S}=0$ and $\delta_h{S}=0$, requires to regard $\delta_v
u^\alpha$, $\delta_v{\pi_\alpha}$ and $\delta x^\mu$ as
independent components and leads to the canonical field equations
\begin{equation}\begin{array}{l}\label{cnleqf}
{\dot u}^{\alpha}(x)=
 \frac{\partial {H}}{\partial \pi_{\alpha
}},\\[2mm]
{\dot \pi}_{\alpha}(x)=-\frac{\partial {H}}{\partial
u^{\alpha}}+\nabla_a (\frac {\partial{H}}{\partial(\nabla_a
u^{\alpha})}),
\end{array}\end{equation}
and an identity between the canonical operators and
conservation property for the energy-momentum tensor
\begin{equation}\label{emcnsvh}
[{H}_{\pi^\alpha}]\partial_\nu \pi^\alpha-[{
H}_{u^\alpha}]\partial_\nu u^\alpha+\partial^\mu
T_{\mu\nu}-\partial_\nu{H}=0.
\end{equation}

It should be mentioned that all remarks in the last subsection can
be made here and the theorem 3 can be established as well.

\section{General Variations for Discrete Field Theory}

We now study  the variation problems for the difference
discrete field theory with variable step-lengths.  For simplicity,
we consider the cases of $1+1$ or $2$ dimensional flat base
manifold, i.e. $X^{1,1}$ or $X^2$ endowed with suitable signature
of the metrics.

Let $L^2$ be a right-angle lattice on $X^{1,1}$ or $X^2$ with
nodes $x_\mu^{(i,j)}=(x_1^i, x_2^j), \mu=1, 2, (i,j)\in Z \times
Z$  and variable step-lengths on 2-directions $x_{\mu}$ to be
determined by discrete variation problems, $N$ be all nodes on
$L^2$.  For a given node with coordinates $x_\mu^{(i,j)}$,
$M_D:=M^{(i,j)}$ be the piece of configuration space with a set of
generic field variables $u^{\alpha}(x_\mu^{(i,j)})=u^{\alpha
(i,j)} \in M_D$ at the node $x_\mu^{(i,j)}$, $T{M^{(i,j)}}$ the
tangent bundle of $M^{(i,j)}$ with the set of field variables and
their differences $(u^{\alpha (i,j)}, \Delta_{\mu}u^{\alpha
(i,j)})  \in T(M^{(i,j)})$, $F(T{M^{(i,j)}})$ the function space
on $T{M^{(i,j)}}$.  Let
 ${\cal N}^{(i,j)}$ be the set of nodes neighboring to
$x_\mu^{(i,j)}$ with index set ${I}^{(i,j)}=Ind({\cal
N}^{(i,j)})$,
a set of nodes related to $x_\mu^{(i,j)}$ by the differences, $
{\cal M}
^{(i,j)}=\bigcup_{Ind({ N})| {I}^{(i,j)}}{ M^{(i,j)}}$ the union
of the pieces of configuration space on ${X}^{(i,j)}$. $F(T({{\cal
M}
^{(i,j)}}))$ function space on $T({{\cal M}
^{(i,j)}})$.

Since $L^2$ is a right-angle lattice, it should have only two
possibilities for the variable step-lengths: either equal
step-length variation along two direction simultaneously while
along each direction the step-lengths are variable, or along one
direction the step-length is fixed while along the other it is
variable.

\subsection{Variable difference Lagrangian field theory}

The difference Lagrangian for a set of the generic fields
$u^{\alpha}, \alpha=1, \cdots, r,$ is a functional
on $F(T({{M}
^{(i,j)}}))$ and suppose to be the first order of
differences of the fields for simplicity%
\begin{equation}\label{lftd}
{{L}_D}^{(i,j)}={L}_D(u^{\alpha (i,j)}, {\Delta_{\mu} u}^{\alpha
(i,j)}, x_\mu^{(i,j)}), \qquad \mu=1,2,%
\end{equation}
where as just mentioned, $x_\mu^{(i,j)}=(x_1^{(i)}, x_2^{(j)})$,
$u^{\alpha (i,j)}=u^\alpha(x_\mu^{(i,j)})$ and
\begin{equation}
{\Delta_1 u}^{\alpha (i,j)}=\frac{u^{\alpha (i+1,j)}-u^{\alpha
(i,j)}}{x_1^{(i+1)}-x_1^{(i)}}, \qquad {\Delta_2 u}^{\alpha
(i,j)}=\frac{u^{\alpha (i,j+1)}-u^{\alpha (i,j)}}{x_2^{(j+1)}-x_2^{(j)}}.%
\end{equation}
The discrete action ${S}_D$ now reads
\begin{equation}\label{sftd}
{S }_D=\sum_{i,j}\Delta_1 x_1^{(i)}\Delta_2 x_2^{(j)}
{{L}_D}^{(i,j)},%
\end{equation}
where $\Delta_1 x_1^{(i)}=x_1^{(i+1)}-x_1^{(i)}, \Delta_2
x_2^{(j)}=x_2^{(j+1)}-x_2^{(j)}$.

 Let us consider the coordinates
of nodes on the lattice are subject to  infinitesimal deformations
that still keep $L^2$ as a right-angle lattice
\begin{equation}\label{dfmtnxd}
x_\mu^{(i,j)}\rightarrow x'_\mu(x_1^{(i,j)},
x_2^{(i,j)})=x_\mu^{(i,j)}+\delta x_\mu^{(i,j)},
\end{equation}
the corresponding changes in the fields are
\begin{equation}\begin{array}{crl}\label{dfmtnud}
u^\alpha (x)^{ (i,j)}&\rightarrow & u'^\alpha (x')^{
(i,j)}=u^\alpha (x)^{ (i,j)}+\delta_tu^\alpha (x)^{ (i,j)},\\
&&\delta_tu^\alpha (x)^{ (i,j)}=\delta_v u^\alpha (x)^{
(i,j)}+\delta_h u^\alpha (x)^{ (i,j)},\\
&&\delta_v u^\alpha (x)^{ (i,j)}:= u'^\alpha (x)^{
(i,j)}- u^\alpha (x)^{ (i,j)},\\
&&\delta_h u^\alpha (x)^{ (i,j)}:=u'^\alpha (x')^{
(i,j)}-u'^\alpha (x)^{ (i,j)}\\
&&\qquad\qquad\quad=u^\alpha (x')^{(i,j)}-u^\alpha (x)^{(i,j)}+O(\delta^2)\\
&&\qquad\qquad\quad=\delta x^{\mu (i,j)}\Delta_\mu u^\alpha (x)^{
(i,j)}.
\end{array}\end{equation}
For the differences of fields, $\Delta_\mu u^\alpha (x)^{ (i,j)}$,
the corresponding changes
\begin{equation}\begin{array}{crl}\label{vrpaud}
\Delta_\mu u^\alpha (x)^{ (i,j)}&\rightarrow & \Delta'_\mu
u'^\alpha (x'(x^{ (i,j)}))=\Delta_\mu u^\alpha (x)^{
(i,j)}+\delta_t\Delta_\mu u^\alpha (x)^{ (i,j)},\\
&&\delta_t\Delta_\mu u^\alpha (x)^{ (i,j)}=\delta_v\Delta_\mu
u^\alpha (x)^{ (i,j)}+\delta_h\Delta_\mu u^\alpha (x)^{ (i,j)},\\
&&\delta_v\Delta_\mu u^\alpha (x)^{ (i,j)}:=\Delta_\mu u'^\alpha
(x)^{ (i,j)}
-\Delta_\mu u^\alpha (x)^{ (i,j)}, \\
&&\delta_h\Delta_\mu u^\alpha (x)^{ (i,j)}:=\Delta'_\mu
u'^\alpha (x')^{ (i,j)}-\Delta_\mu u'^\alpha (x)^{ (i,j)}\\
&&\qquad\qquad\qquad\quad=\Delta_\mu u^\alpha (x')^{
(i,j)}-\Delta_\mu u^\alpha (x)^{ (i,j)}+o(\delta^2),
\end{array}\end{equation}
 can be calculated to get
\begin{equation}\begin{array}{crl}\label{vrpaud2}
\delta_v\Delta_\mu u^\alpha (x)^{ (i,j)}&=&\Delta_\mu (\delta_v
u^\alpha (x)^{ (i,j)})\\
 \delta_t\Delta_\mu u^\alpha (x)^{ (i,j)}&=&\Delta_\mu \delta_tu^\alpha (x)^{ (i,j)}
 -\Delta_\mu \delta x^{\nu (i,j)}\cdot \Delta_\nu u^\alpha (x)^{
 (i,j)}.
\end{array}\end{equation}
Using the Leibniz law (\ref{lbnz}) for differences in each
direction, it follows
\begin{equation}\begin{array}{crl}\label{vrpaud3}
\delta_t\Delta_1 u^\alpha (x)^{ (i,j)}&=&\Delta_1 \delta_v
u^\alpha (x)^{ (i,j)}+\delta x^{\nu (i+1,j)}\cdot\Delta_1
\Delta_\nu u^\alpha (x)^{ (i,j)},\\
\delta_t\Delta_2 u^\alpha (x)^{ (i,j)}&=&\Delta_2 \delta_v
u^\alpha (x)^{ (i,j)}+\delta x^{\nu (i,j+1)}\cdot\Delta_2
\Delta_\nu u^\alpha (x)^{ (i,j)}.\\
\end{array}\end{equation}

The total variation of discrete action (\ref{sftd}) can be
calculated
\begin{equation}\begin{array}{crl}\label{vrsd1}
\delta_t{S }_D&=&\sum_{i,j}\Delta_1x_1^{(i)} \cdot \Delta_2 x_2^{(j)}%
(\Delta_1 \delta x_1^i {{L}_D}^{(i,j)}+\Delta_2 \delta x_2^j
{{L}_D}^{(i,j)}+\delta_t{{L}_D}^{(i,j)}).%
\end{array}\end{equation}
Using formulae
\begin{equation}\begin{array}{crl}\label{vractn2}
\Delta_1 \delta x_1^i {{L}_D}^{(i,j)}&=&\Delta_1(\delta x_1^i
{{L}_D}^{(i-1,j)})-\delta x_1^i \Delta_1{{\cal
L}_D}^{(i-1,j)},\\
\Delta_2 \delta x_2^j {{L}_D}^{(i,j)}&=&\Delta_2(\delta x_2^j
{{L}_D}^{(i,j-1)})-\delta x_2^j \Delta_2{{\cal L}_D}^{(i,j-1)},
\end{array}\end{equation}
and
\begin{equation}\begin{array}{crl}\label{vrlg}
\delta_t{{L}_D}^{(i,j)}&=&\frac{\partial{{\cal
L}_D}^{(i,j)}}{\partial u^{\alpha(i,j)}}\delta_t
u^{\alpha(i,j)}+\frac{\partial{{L}_D}^{(i,j)}}{\partial
(\Delta_\mu u^{\alpha(i,j)}) }\delta_t\Delta_\mu
u^{\alpha(i,j)}+\frac{\partial{{L}_D}^{(i,j)}}{\partial
x^{\mu (i,j)}}\delta x^{\mu (i,j)} \\%
&=&\frac{\partial{{L}_D}^{(i,j)}}{\partial
u^{\alpha(i,j)}}\delta_tu^{\alpha(i,j)}+ \frac{\partial{{
L}_D}^{(i,j)}}{\partial (\Delta_1 u^{\alpha(i,j)})}(\Delta_1
\delta_t
u^{\alpha(i,j)}-(\Delta_1\delta x^{\mu (i,j)})\cdot \Delta_\mu u^{\alpha(i,j)}) \\
&+&\frac{\partial{{L}_D}^{(i,j)}}{\partial (\Delta_2
u^{\alpha(i,j)})}(\Delta_2 \delta_t
u^{\alpha(i,j)}-(\Delta_2\delta x^{\mu (i,j)})\cdot \Delta_\mu
u^{\alpha(i,j)})+\frac{\partial{{L}_D}^{(i,j)}}{\partial
x^{\mu (i,j)}}\delta x^{\mu (i,j)}\\
&=&[{L}_{u^{\alpha (i,j)}}]\delta_t
u^{\alpha(i,j)}\\%
&+&\Delta_1(\frac{\partial{{L}_D}^{(i-1,j)}}{\partial (\Delta_1
u^{\alpha(i-1,j)})} \delta_t
u^{\alpha(i,j)}-\frac{\partial{{L}_D}^{(i-1,j)}}{\partial
(\Delta_1 u^{\alpha(i-1,j)})}\Delta_\mu u^{\alpha(i-1,j)}\delta x^{\mu (i,j)})\\%
&+&\Delta_2(\frac{\partial{{L}_D}^{(i,j-1)}}{\partial (\Delta_2
u^{\alpha(i,j-1)})} \delta_t
u^{\alpha(i,j)}-\frac{\partial{{L}_D}^{(i,j-1)}}{\partial
(\Delta_2 u^{\alpha(i,j-1)})}\Delta_\mu u^{\alpha(i,j-1)}\delta
x^{\mu (i,j)})\\
&+&\Delta_1(\frac{\partial{{L}_D}^{(i-1,j)}}{\partial (\Delta_1
u^{\alpha(i-1,j)})}\Delta_\mu u^{\alpha(i-1,j)})\delta
x^{\mu (i,j)}\\
&+&\Delta_2(\frac{\partial{{L}_D}^{(i,j-1)}}{\partial (\Delta_2
u^{\alpha(i,j-1)})}\Delta_\mu
u^{\alpha(i,j-1)})\delta x^{\mu (i,j)}\\
&+&\frac{\partial{{L}_D}^{(i,j)}}{\partial x^{\mu (i,j)}}\delta
x^{\mu (i,j)},
\end{array}\end{equation}
we get
\begin{equation}\begin{array}{crl}\label{vrsd3}
\delta_t{S }_D &=&\delta_v{S }_D+\delta_h{S
}_D\\&=&\sum_{i,j}\Delta_1 x_1^{(i)} \cdot \Delta_2 x_2^{(j)}
\{[{L}_{u^{\alpha (i,j)}}]\delta_tu^{\alpha(i,j)}\\%
&+&\Delta_1 (\frac{\partial{{L}_D}^{(i-1,j)}}{\partial (\Delta_1
u^{\alpha(i-1,j)})} \delta_t
u^{\alpha(i,j)}-(\frac{\partial{{L}_D}^{(i-1,j)}}{\partial
(\Delta_1 u^{\alpha(i-1,j)})}\Delta_\mu
u^{\alpha(i-1,j)}-\delta_{1\mu} {{L}_D}^{(i-1,j)})\delta
x^{\mu (i,j)})\\%
&+&\Delta_2(\frac{\partial{{L}_D}^{(i,j-1)}}{\partial (\Delta_2
u^{\alpha(i,j-1)})} \delta_t
u^{\alpha(i,j)}-(\frac{\partial{{L}_D}^{(i,j-1)}}{\partial
(\Delta_2 u^{\alpha(i,j-1)})}\Delta_\mu u^{\alpha(i,j-1)}
-\delta_{2\mu} {{L}_D}^{(i,j-1)})\delta
x^{\mu (i,j)})\\
&+&\Delta_1(\frac{\partial{{L}_D}^{(i-1,j)}}{\partial (\Delta_1
u^{\alpha(i-1,j)})}\Delta_\mu u^{\alpha(i-1,j)}-\delta_{1\mu}
{{L}_D}^{(i-1,j)})\delta
x^{\mu (i,j)}\\
&+&\Delta_2(\frac{\partial{{L}_D}^{(i,j-1)}}{\partial (\Delta_2
u^{\alpha(i,j-1)})}\Delta_\mu
u^{\alpha(i,j-1)}-\delta_{2\mu} {{L}_D}^{(i,j-1)})\delta x^{\mu (i,j)}\\
&+&\frac{\partial{{L}_D}^{(i,j)}}{\partial x^{\mu
(i,j)}}\delta x^{\mu (i,j)}\} \\%
&=&\sum_{i,j}\Delta_1 x_1^{(i)} \cdot \Delta_2 x_2^{(j)}
\{[{L}_{u^{\alpha (i,j)}}]\delta_tu^{\alpha(i,j)}\\%
&+&\Sigma_{\mu,\nu=1,2}\Delta_\mu(\frac{\partial{{
L}_D}^{(i,j)}}{\partial (\Delta_\mu u^{(i,j)})}\delta_t
u^{(i,j)}-E_\mu^{-1}{{
T}_D}_\nu ^{\mu (i,j)}\delta x^{\nu (i,j)})\\
&+&\Sigma_{\nu,\mu=1,2}(\Delta_\nu {E_{\nu}}^{-1}{{{
T}_D}_\mu^{\nu}}^{(i,j)}+\frac{\partial{{ L}_D}^{(i,j)}}{\partial
x^{\mu (i,j)}})\delta x^{\mu (i,j)}\},
\end{array}\end{equation}
where $E_\mu, \mu=1,2$, $[{L}_{u^{\alpha (i,j)}}]$ and ${{T}_{D
\mu \nu}} ^{ (i,j)}$ are shift operators, discrete Euler-Lagrange
operator and energy-momentum tensor respectively
\begin{equation}\begin{array}{crl}
\label{shfts}
E_1f^{(i,j)}=f^{(i+1,j)},\quad E_1^{-1}f^{(i,j)}=f^{(i-1,j)},\\%
E_2f^{(i,j)}=f^{(i,j+1)},\quad E_2^{-1}f^{(i,j)}=f^{(i,j-1)};%
\end{array}\end{equation}%
\begin{equation}\label{elftd}
 [{L}_{u^{\alpha (i,j)}}]:= \frac
{\partial {{L}_D}^{(i,j)}} {\partial u^{\alpha (i,j)}} -\Delta_1
(\frac {\partial {{L}_D}^{(i-1,j)}} {\partial (\Delta_1 u^{\alpha
(i-1,j)})}) -\Delta_2(\frac {\partial {{
L}_D}^{(i,j-1)}} {\partial (\Delta_2 u^{\alpha (i,j-1)})});%
\end{equation}
\begin{equation}\label{emftd}
{{T}_{D \mu \nu}} ^{ (i,j)}:=\frac{\partial {{ L}_D}^{(i,j)}
}{\partial (\Delta^\mu u^{\alpha (i,j)})}\Delta_\nu u^{\alpha
(i,j)} - {{L}_D}^{(i,j)}\eta_{\mu\nu}.
\end{equation}

Regarding $\delta_v u^{\alpha(i,j)}$ and $\delta x^{\nu (i,j)}$
are independent variational bases, $\delta_t{S }_D=0$, or
$\delta_v {S }_D=0$ and $\delta_h {S }_D=0$, lead to the discrete
Euler-Lagrange equation
\begin{equation}\label{eleqd1}
\frac {\partial {{L}_D}^{(i,j)}} {\partial u^{\alpha (i,j)}}
-\Delta_1 (\frac {\partial {{L}_D}^{(i-1,j)}} {\partial (\Delta_1
u^{\alpha (i-1,j)})}) -\Delta_2(\frac {\partial {{
L}_D}^{(i,j-1)}} {\partial (\Delta_2 u^{\alpha (i,j-1)})})=0,
\end{equation}
a relation between the Euler-Lagrange operator and the
(difference) divergence of the discrete energy-momentum tensor
that may determine the step-lengths
\begin{equation}\label{el-emt}
[{L}_{u^{\alpha (i,j)}}]\Delta_\nu u^\alpha (x)^{
(i,j)}+\Sigma_{\mu=1,2}\{ \Delta_\mu E_\mu^{-1}{{T}_D}_\nu ^{\mu
(i,j)}+\frac{\partial{{L}_D}^{(i,j)}}{\partial x^{\nu (i,j)}}\}=0.
\end{equation}

It is obvious that all these discrete equation, relation and
properties have correct continuous limits respectively.
Furthermore, due to the discrete Lagrangian (\ref{lftd}) depends
on the differences explicitly, it is possible to introduce the
discrete canonical momentum and discrete Legendre transformation
to transfer to the discrete Hamiltonian formalism as will be shown
in next subsection.

\vskip 3mm {\bf Remark 5.1:}

We may introduce exterior differential operators $\hat{d}$, $d_v$
and $\hat{d}_h$ on $T^*(M \times X_D)$, $T^*M$ and $T^* X_D$
respectively. They are nilpotent and satisfy
\begin{equation}\label{dhatf}
\hat{d}=d_v+\hat{d}_h,\qquad \{d_v, \hat{d}_h\}=0.
\end{equation}
Especially, $\hat{d}_h$ is due to the difference on $X_D$ and
satisfy  Leibniz's law for ordinary forms {\footnote{It is needed
some noncommutative differential calculus to completely clarify
the properties of $\hat{d}_h$. For the case that $\Delta x^\mu$
are fixed, this noncommutative differential calculus can be found
in \cite{gwwww},\cite{gwz}. For the case of variable step-lengths,
similar noncommutative differential calculus can be established
\cite{gw}.}}.

\vskip 3mm {\bf Remark 5.2:}

Actually, analog to the case with fixed  step-lengths \cite{glw},
\cite{glww}, it can be established the difference version for the
Euler-Lagrange cohomology and the necessary and sufficient
condition for the difference conservation law of the discrete
multi-symplectic 2-forms.

From $\delta_v S_D$ in (\ref{vrsd3}), it is easy to see that we
may take $d_v$ on $S_D$ to get
\begin{equation}\label{dSDf}
d_v S_D=\sum_{(i,j)} \Delta_1 x_1^i \Delta_2 x_2^j d_v
{L_D}^{(i,j)}, \qquad d_v {L_D}^{(i,j)}={{\cal
E}_D}^{(i,j)}+\Delta_\mu
{\theta_D}^{\mu (i,j)},
\end{equation}
where ${{\cal E}_D}^{(i,j)}$, ${\theta_D}^{\mu (i,j)}, \mu=1, 2$
are the discrete Euler-Lagrange 1-form and multi-symplectic
potential 1-forms respectively
\begin{eqnarray}
{{\cal E}_D}^{(i,j)}&:=&[L_{D u^{\alpha (i,j)}}]d_v u^{\alpha (i,j)},\\
{\theta_D}^{1(i,j)}:=\frac {\partial { L}_D^{(i-1,j)}} {\partial
({\Delta_1 u }^{\alpha (k-1,l)})} du^{\alpha (k,l)}, && {\theta_{D
L}}^{2 (i,j)}:=\frac {\partial { L}_D^{(i,j-1)}} {\partial
({\Delta_2 u}^{\alpha (k,l-1)})}
du^{\alpha (k,l)}.%
\end{eqnarray}
Then due to the nilpotency of $d_v$, it is straightforward to get
\begin{equation}\label{ddsDf}
d_v{{\cal E}_D}^{(i,j)}+\Delta_\mu {\omega_D}^{\mu (i,j)}=0,\qquad
{\omega_D}^{\mu
(i,j)}:=d_v{\theta_D}^{\mu(i,j)}.
\end{equation}
Therefore, we may get the discrete version for the theorem 3
\cite{glw}, \cite{glww}:

\vskip 3mm{\it Theorem 4: For all discrete Lagrangian of a kind of
discrete field theories with first order of differences on the
bundle $E(X_D, Q, \pi) \simeq M \times X_D$,

1. The following discrete version of the Euler-Lagrange cohomology
is nontrivial:

{\centerline{$H_{DFT}$:=\{Closed Euler-Lagrange forms\}/ \{Exact
Euler-Lagrange forms\}.}}

2. The necessary and sufficient condition for conservation of the
discrete multi-symplectic 2-forms, i.e.
\begin{equation}\label{csvmsyD}
\Delta_\mu{\omega_D}^{\mu (i,j)}=0,
\end{equation}
is the corresponding discrete Euler-Lagrange 1-form being closed.
}


\vskip 3mm {\bf Remark 5.3:}

In this paper, $L^2$ is an infinite lattice. It is reasonable to
consider a finite lattice. We will publish the issues on this
topic elsewhere.

\subsection{Variable difference Hamiltonian field theory}
Consider $X^{(1,1)}$, on which there is a right-angle
lattice $L^2$ with variable step-lengths in each direction, is the
base space.

We first define a set of the discrete canonical conjugate momenta
on the tangent space $T({\cal M}
^{(i,j)})$ of $ {\cal  M}
^{(i,j)}=\bigcup_{Ind({N}) | { I}^{(i,j)}}{ M^{(i,j)}}$ the union
of the pieces of configuration space on ${X}^{(i,j)}$, which are
the set of nodes neighboring to the node $x_\mu^{(i,j)}$:
\begin{equation}\label{cmftd}%
 {\pi_{\alpha}}^{
(i,j)}=\frac {\partial {{L}_D}^{(i-1,j)}} {\partial ({\Delta_t
{u}^{\alpha
(i-1,j)})}}.%
\end{equation}
 The difference Hamiltonian is introduced through
the discrete Legendre transformation
\begin{equation}\label{ltfmfthd}
{{H}_D}^{(i,j)}({u}^{\alpha (i,j)}, \pi_{\alpha}^{(i+1,j)};
x^{(i,j)}) ={\pi_{\alpha}}^{(i+1,j)}{\Delta_t {u}^{\alpha
(i,j)}}-{{L}_D}^{(i,j)}.
\end{equation}

The action functional (\ref{sftd}) now is expressed as
\begin{equation}\label{sftd2}
{S}_D=\sum_{(i,j) \in {Z \times Z}}\Delta_\mu x^{\mu
(i,j)}({\pi_{\alpha}}^{(i+1,j)}{\Delta_t {u}^{\alpha
(i,j)}}-{{H}_D}^{(i,j)}).
\end{equation}
The total variation of the action $\delta_t{S}_D$ can be
calculated and separated into two parts, i.e. the vertical
variation $\delta_v{S}_D$ and the horizontal variation
$\delta_h{S}_D$
\begin{equation}\label{voad2}
\delta_t{S}_D =\delta_v{S}_D +\delta_h{S}_D,
\end{equation}
\begin{equation}\label{voad3}\begin{array}{rcl}
\delta_v{S}_D &=&\sum_{(i,j) \in {Z \times Z}}\Delta_\mu x^{\mu
(i,j)} \{\delta_v{\pi_{\alpha}}^{(i+1,j)}[{
H}_{{\pi_{\alpha}}^{(i+1,j)}}] -[{H}_{u^{\alpha (i,j)}}]
\delta_vu^{\alpha (i,j)}\\
&-&\Sigma_{\mu,\nu=1,2}\Delta_\mu(\frac{\partial{{
H}_D}^{(i,j)}}{\partial (\Delta_\mu u^{(i,j)})}\delta_v u^{(i,j)})
\},
\end{array}\end{equation}
\begin{equation}\label{voad4}\begin{array}{rcl}
\delta_h{S}_D&=& \sum_{(i,j) \in {Z \times Z}}\Delta_\mu x^{\mu
(i,j)} \{\delta_h{\pi_{\alpha}}^{(i+1,j)}[{
H}_{{\pi_{\alpha}}^{(i+1,j)}}] -[{H}_{u^{\alpha (i,j)}}]
\delta_hu^{\alpha (i,j)}\\
&-&\Sigma_{\mu,\nu=1,2}\Delta_\mu(\frac{\partial{{
H}_D}^{(i,j)}}{\partial (\Delta_\mu u^{(i,j)})}\delta_h
u^{(i,j)}+E_\mu^{-1}{{
T}_D}_\nu ^{\mu (i,j)}\delta x^{\nu (i,j)})\\
&+&\Sigma_{\nu,\mu=1,2}(\Delta_\nu {E_{\nu}}^{-1}{{{
T}_D}_\mu^{\nu}}^{(i,j)}-\frac{\partial{{ H}_D}^{(i,j)}}{\partial
x^{\mu (i,j)}})\delta x^{\mu (i,j)}\}. \end{array}\end{equation}
Here
\begin{equation}\begin{array}{rcl}\label{cop}
[{H}_{{\pi_{\alpha}}^{(i+1,j)}}]&:=&\Delta_t u^{\alpha
(i,j)}-\frac{\partial{{H}_D}^{(i,j)}}{\partial
{\pi_{\alpha}}^{(i+1,j)}},\\[1mm]
[{H }_{u^{\alpha (i,j)}}] &:=&\Delta_t {\pi_{\alpha}}^{(i,j)}
+\frac {\partial {H }_D^{(i,j)}} {\partial u^{\alpha (i,j)}}
-\Delta_x(\frac {\partial {H }_D^{(i,j-1)}} {\partial (\Delta_x
u^{\alpha (i,j-1)})}).
\end{array}%
\end{equation}

Regarding $\delta_v u^{\alpha(i,j)}$,
$\delta_v{\pi_{\alpha}}^{(i+1,j)}$ and $\delta x^{\nu (i,j)}$ are
independent variational bases,  $\delta_v {S }_D=0$ due to
discrete Hamilton's principle and $\delta_h {S }_D=0$ due to
discreteized re-parameterization invariance on $L^2$, i.e.
$\delta_t{S }_D=0$, lead to the discrete canonical field equations
\begin{equation}\begin{array}{rcl}
\Delta_t u^{\alpha (i,j)}&=&\frac{\partial{{
H}_D}^{(i,j)}}{\partial
{\pi_{\alpha}}^{(i+1,j)}},\\
\Delta_t {\pi_{\alpha}}^{(i,j)}&=&-\frac {\partial {H }_D^{(i,j)}}
{\partial u^{\alpha (i,j)}} +\Delta_x(\frac {\partial {H
}_D^{(i,j-1)}} {\partial (\Delta_x u^{\alpha (i,j-1)})}),
\end{array}\end{equation}
the canonical form of the relation (\ref{el-emt}) that may
determine the step-lengths
\begin{eqnarray}
[{H}_{{\pi_{\alpha}}^{(i+1,j)}}]\Delta_\nu{\pi_{\alpha}}^{(i+1,j)}
-[{H}_{u^{\alpha (i,j)}}] \Delta_\nu u^{\alpha (i,j)}
+\Sigma_{\mu=1,2}\{ \Delta_\mu E_\mu^{-1}{{T}_D}_\nu ^{\mu
(i,j)}-\frac{\partial{{H}_D}^{(i,j)}}{\partial x^{\nu (i,j)}}\}=0.
\end{eqnarray}

It should also be mentioned that all remarks in the last
subsection can be made hare and the theorem 4 can also be
established in this discrete Hamiltonian formalism for field
theory.

\section{Concluding remarks}

In this paper, the variable difference variational approach with
variable step-lengths has been proposed. It is a generalized
version of the difference variational approach with fixed
step-lengths proposed in \cite{glw}, \cite{glww}. The approach has
been applied to both Lagrangian and Hamiltonian formalism for
discrete mechanics and field theory. Although what have been dealt
with are the systems with first order differences, the key points
are available for the systems with higher order differences.
Obviously, both approaches are different from either Lee's
discrete variation with variable time-steps \cite{td1},
\cite{td2}, \cite{td3} or Veselov's one with fixed time-steps for
the discrete classical mechanics \cite{av88}, \cite{MV91}. They
are also different from the discrete variation approach to field
theory in \cite{MPS98} that is a generalization of Veselov's
approach. In our approaches the differences with either variable
step-lengths or the fixed ones are regarded as discrete
derivatives being entire geometric objects. This is more obvious
and natural from the viewpoint of noncommutative geometry and more
analogical to the continuous mechanics and field theory.
Therefore, in the continuous limit, the results given here by the
variable difference  variational approach unaffectedly lead to the
correct continuous counterparts not only for the equations of
motion and symplectic or multisymplectic preserving properties,
but also for the conservation laws, especially for the energy
conservation.

In view of the structure-preserving criterion for the discrete
systems, there are more advantages for the variable difference
discrete variational approach. Eventually, this has been already
seen in \cite{cgw} where in taking the continuous limits for the
discrete variation problems, which is a generalized version of
Lee-Veselov's variation, combining first discrete objects into
some difference form is more controllable.

With variable step-lengths it is, of course, more or less
straightforward to generalize the symplectic and multisyplectic
schemes as  ones that are not only symplectic and multisymplectic
preserving but also discretely energy conserved as has been done
for variational symplectic energy-momentum integrators in discrete
Lagrangian formalism \cite{kmo99}, \cite{cgw}, and in discrete
Hamiltonian formalism \cite{cgw1}. But, these discrete formalisms
do not transform to each other via discrete Legendre
transformation.

The difference  variational approach has been applied to the
symplectic algorithm and multisymplectic one for both Lagrangian
and Hamiltonian formalism in \cite{glww}. It has been shown that
the discrete integrants can be combined together in certain manner
as certain geometric objects in order to construct some numerical
schemes with fixed step-lengths as variational integrators such as
the midpoint scheme in symplectic algorithm and the Preissman
scheme in multisymplectic algorithm. Obviously, the variable
difference variational approach should be able to apply to the
symplectic  and multisymplectic algorithms with variable
step-lengths for both Lagrangian and Hamiltonian formalism. This
issue will be published in details elsewhere.

It has been found that the necessary and sufficient conditions for
the symplectic 2-form preserving in mechanics and the
multisymplectic 2-forms preserving in field theories are the
corresponding Euler-Lagrange 1-form is closed in the relevant
Euler-Lagrange cohomology \cite{glw}, \cite{glww1}. This is also
true for the discrete cases \cite{glw}, \cite{glww} as well as the
symplectic and multisymplectic algorithms \cite{glww}. For the
cases studied in this paper, the Euler-Lagrange cohomology should
also be true for the various variation problems with variable
domains or step-lengths. In fact, this matter is already indicated
by the boundary terms in the vertical parts of these variation
problems. We will explain this issue in more detail elsewhere.

 We have almost completely employed the ordinary description in a
coordinate manner in this paper not only in order to be more
easily understood, especially for non-mathematician, but for
dealing with both continuous and discrete cases in an analogical
manner. It should be mentioned however that both the variation
problems and the Euler-Lagrange cohomology for continuous cases
could be dealt with in a coordinate free version in terms of jet
bundle and variational bicomplex (see, for example, \cite{adsn89},
\cite{adsn92}). Although as far as the local issues are concerned,
the essentials are almost the same. The coordinate free expression
should be able to contain more general and complicated cases with
nontrivial topology. On the other hand, however, for the discrete
cases the ordinary jet bundle and variational bicomplex approach
should be generalized to the ones include non-commutative
differential calculus in principle. We will present the
variational bicomplex approach to the issues in this paper
elsewhere, specially the one with non-commutative differential
calculus to the discrete cases \cite{gw}.

\vskip 8mm

\centerline{\bf Acknowledgments}

We would like to thank Profs/Drs/Mr/Ms J.B. Chen, Y.Q. Li, T.T. Liu,
 M.Z. Qin,  Z.J. Shang,
G. Sun, Y.J. Sun, Y.F. Tan, Y.S. Wang, X.N. Wu and Z. Xu for
valuable discussions. This work was supported in part by the
National Natural Science Foundation of China and the National Key
Project  for Basic Research of China G1998030601.

\end{document}